\documentclass[fleqn,usenatbib]{mnras}

\usepackage{newtxtext,newtxmath}

\usepackage[T1]{fontenc}
\usepackage{soul}

\DeclareRobustCommand{\VAN}[3]{#2}
\let\VANthebibliography\thebibliography
\def\thebibliography{\DeclareRobustCommand{\VAN}[3]{##3}\VANthebibliography}

\usepackage{graphicx}	
\usepackage{amsmath}	
\usepackage{comment}
\allowdisplaybreaks
\title[Dynamical tides in stratified rapid rotators]
{Dynamical tides in Jupiter and other rotationally flattened planets and stars with stable stratification}

\author[Dewberry]{
Janosz W. Dewberry,$^{1}$\thanks{E-mail: jdewberry@cita.utoronto.ca}
\\
$^{1}$ Canadian Institute for Theoretical Astrophysics, 60 St. George Street, Toronto, ON M5S 3H8, Canada
}

\date{Accepted XXX. Received YYY; in original form ZZZ}

\pubyear{2022}

\begin{document}
\label{firstpage}
\pagerange{\pageref{firstpage}--\pageref{lastpage}}
\maketitle

\begin{abstract}
We develop a numerical method for directly computing the dissipative dynamical tidal response of rapidly rotating, oblate stars and gaseous planets with realistic internal structures. Applying these calculations to neutrally and stably stratified polytropes, we identify the most relevant resonances in models with rotation rates up to nearly the mass-shedding limit. We then compute the dynamical tidal response for Jupiter interior models including both stably stratified and convective regions. These calculations show that resonances involving mixed waves with \emph{both} gravito-inertial and purely inertial character are capable of explaining a discrepancy between observations and hydrostatic calculations of Jupiter's response to tidal forcing by Io. This result contrasts with recent work that excluded Jupiter's rotational flattening, and opens the door to resonances involving a wider range of internal oscillation modes than previously considered.
\end{abstract}




\section{Introduction}
Tidal interactions likely play a role in a wide variety of astrophysical scenarios, mediating the interactions and influencing the  orbital evolution of moons, planets, stars, and compact objects alike \citep{Ogilvie2014}. Despite their wide-reaching relevance, several quantitative details of tidal exchanges in energy and angular momentum have proven difficult to square with observations in both astrophysics and planetary sciences, particularly in situations where one or more of the tidally interacting bodies is rotating. Setting aside complications related to the Coriolis force \citep[e.g.,][]{Ogilvie2004,Ogilvie2009,Ogilvie2013,Wu2005,Wu2005b,Ivanov2007,Goodman2009,Rieutord2010,Lin2021}, relatively little has been done to characterize the effects that changes in stellar and planetary shape due to rotation have on dynamical (i.e., frequency-dependent) tidal distortion and dissipation. 

The gas giant planets in our own solar system motivate such characterization. Jupiter and Saturn respectively rotate at nearly $30\%$ and $40\%$ of their break-up angular velocities, and are consequently oblate. Moreover, measurements of the shape of the tidal bulge raised on Jupiter by Io---characterized by so-called ``Love numbers'' \citep{Durante2020}---deviate significantly from theoretical predictions for purely static tidal perturbers \citep{Wahl2017,Wahl2020,Nettelmann2019}. This discrepancy has inspired the suggestion that Io's orbit may be in resonance with the natural frequency of an internal oscillation mode (in particular a gravito-inertial mode) of Jupiter \citep{Idini2022b}. However, recent calculations \citep{Lin2023} have cast doubt on the ability of such a resonance to reconcile hydrostatic calculations with the observations. Notably, both \citet{Idini2022b} and \citet{Lin2023} excluded the effects of centrifugal flattening in their calculations of tidally driven oscillations. 

We use spectral methods to directly compute the dissipative tidal response of rapidly rotating and centrifugally flattened planets and stars. Our numerical method is valid for arbitrarily rapid and differential rotation on cylinders, incorporating dissipation via a viscous stress that self-consistently includes rotational flattening. Limiting our focus to rigid rotation in this work, we first apply this method to computing the frequency-dependent, dynamical tidal response of $\gamma=5/3$ and $3/2$ polytropes rotating at up to nearly the mass-shedding limit. We then compute the tidal response for Jupiter interior models that include both stably stratified and convective regions. The latter calculations demonstrate that resonant wave excitation by dynamical tides is in fact capable of reconciling the discrepancy between observations and hydrostatic calculations of Jupiter and Io's interaction, but only if the non-spherical aspects of Jupiter's rotation are accounted for. Our calculations further suggest that a wider set of internal oscillations than considered by \citet{Idini2022b} should make viable candidates for a Jupiter-Io resonance. 

This paper is structured as follows. Section \ref{sec:methods} introduces our numerical method, and covers relevant background information. Although many of the technical details may be skipped by those interested only in our results, we note that subsection \ref{sec:love} lays out conventions for Love number definitions that are important to interpreting our calculations. Section \ref{sec:poly} then describes our results for very rapidly rotating polytropes, and Section \ref{sec:Jup} describes our calculations for Jupiter. We conclude in Section \ref{sec:conc}.

\section{Methods and background}\label{sec:methods}
\subsection{Fluid dynamics}
This subsection introduces the equations governing small-amplitude perturbations to oblate gaseous bodies, and our numerical methods for solving them.

\subsubsection{Basic equations}
The Newtonian equation of motion for a self-gravitating fluid with pressure $P$, density $\rho$, gravitational potential $\Phi$, and  velocity ${\bf u}$ is
\begin{equation}\label{eq:EoM0}
    \dfrac{D{\bf u}}{Dt}
    =-\frac{\nabla P}{\rho}
    -\nabla\Phi
    +{\bf F},
\end{equation}
where $D/D_t=\partial_t+{\bf u}\cdot\nabla$ is the convective derivative, and ${\bf F}$ comprises any additional forces. For the case of a viscous fluid subject to a perturbing potential $U$,
\begin{equation}\label{eq:F}
    {\bf F}=-\nabla U + \frac{1}{\rho}\nabla\cdot{\bf T},
\end{equation}
where 
\begin{equation}\label{eq:T}
    {\bf T}=\mu_v\left[
        \nabla{\bf u} 
        +(\nabla{\bf u})^T
        -\frac{2}{3}(\nabla\cdot{\bf u}){\bf I}
    \right]
\end{equation}
is the viscous stress tensor associated with dynamic viscosity $\mu_v$. 

\autoref{eq:EoM0} must be considered simultaneously with the equation of mass conservation,
\begin{equation}\label{eq:cty}
    \dfrac{D\rho}{Dt}
    =-\rho\nabla\cdot{\bf u},
\end{equation}
Poisson's equation
\begin{equation}\label{eq:Poi}
    \nabla^2\Phi=4\pi G\rho
\end{equation}
(here $G$ is the gravitational constant), an equation of state, and the thermal energy equation. Ignoring non-adiabatic heating by viscous dissipation, or cooling by radiation, the latter is given by
\begin{equation}\label{eq:Therm}
    \dfrac{D P}{Dt}
    =-\Gamma_1 P\nabla\cdot{\bf u},
\end{equation}
where $\Gamma_1$ is the first adiabatic exponent. 

\subsubsection{Equilibrium state}
To model the steady state of rotating stars and gaseous planets, we construct axisymmetric, time-independent solutions of Equations \eqref{eq:EoM0}-\eqref{eq:Therm} with equilibrium pressure $P_0,$ density $\rho_0,$ gravitational potential $\Phi_0$ and velocity field ${\bf u}_0=\boldsymbol{\Omega}\times{\bf r}=R\Omega(R)\hat{\boldsymbol{\phi}}$. Here $\boldsymbol{\Omega}$ is an angular velocity that we allow to depend on cylindrical $R=r\sin\theta$ (the distance from the rotation axis). Ignoring ${\bf F}$, such equilibria satisfy
\begin{equation}
    {\bf G}
    =R\Omega^2\hat{\bf R}
    -\nabla \Phi_0,
\end{equation}
where ${\bf G}=\rho_0^{-1}\nabla P_0$ is an effective gravity that includes centrifugal flattening due to rotation. The equilibrium model of the rotating planet or star provides a natural scale for non-dimensionalization: throughout, we adopt units scaled by the total mass and equatorial radius $R_\text{eq}$ (i.e., $G=M=R_\text{eq}=1$). The relevant time-scale is then dictated by the dynamical frequency $\Omega_d=(GM/R_\text{eq}^3)^{1/2}.$

The primary difficulty in computing rotating stellar and planetary equilibria derives from the fact that the oblate, rotationally flattened surface is not known ahead of time for any but the simplest cases. We use the approach to this free-boundary value problem described in \citet{Dewberry2022b} to compute the polytropic models considered in this work. Note that in combination with stable stratification, such rotation can give rise to baroclinic flows involving meridional circulation \citep{Rieutord2006}, which we neglect.

\subsubsection{Linearized equations}
We write $P=P_0+\delta P$, $\rho=\rho_0+\delta\rho$, $\Phi=\Phi_0+\delta\Phi,$ ${\bf u}={\bf u}_0+{\bf v}$, where $\delta P,\delta\rho,\delta\Phi,{\bf v}$ are small-amplitude Eulerian perturbations with a harmonic dependence $\propto\exp[\text{i}(m\phi-\sigma t)]$ on inertial-frame frequency $\sigma$ and azimuthal wavenumber $m.$ Assuming an adiabatic relationship between Lagrangian pressure and density perturbations \citep{LyndenBell1967}, the fluid dynamic equations can then be linearized to find
\begin{align}\label{eq:EoMl}
    -\text{i}\sigma{\bf v}
    +{\bf v}\cdot\nabla{\bf u}_0
    +{\bf u}_0\cdot\nabla{\bf v}
    -{\bf G}\beta
    +(\nabla +\nabla\ln\rho_0)h
    +\nabla\delta\Phi
    &\\\notag
    -\frac{1}{\rho_0}\nabla \cdot {\bf \delta T}
    &=-\nabla U,
\\\label{eq:Tl}
    \delta{\bf T}
    -\mu_v[
    \nabla{\bf v} 
    +(\nabla{\bf v})^T
    -(2/3)(\nabla\cdot{\bf v}){\bf I}
    ]&=0,
\\\label{eq:Ctyl}
    -\text{i}\omega\beta 
    +\frac{1}{\rho_0}\nabla\cdot(\rho_0{\bf v})&=0,
\\\label{eq:TEl}
    -\text{i}\omega (h-c_A^2\beta)
    +({\bf G}-c_A^2\nabla\ln\rho_0)\cdot {\bf v}&=0,
\\\label{eq:Poil}
    4\pi G\rho_0\beta-\nabla^2\delta\Phi&=0.
\end{align}
Here $h=\delta P/\rho_0,$ $\beta=\delta\rho/\rho_0,$ $c_A^2=\Gamma_1P_0/\rho_0,$ and $\omega=\sigma - m\Omega$. For a rigidly rotating body with constant $\Omega$, $\omega$ gives the frequency in the corotating frame. 

Equations \eqref{eq:EoMl}-\eqref{eq:Poil} can be treated as both an inhomogeneous boundary value problem with $\sigma$ and $U$ specified, and an eigenvalue problem with $U\equiv0$ and $\sigma$ unknown. As described in Appendices \ref{app:lin} and \ref{app:num}, we use spectral methods to solve both. This process is complicated by the influence of the Coriolis force (which intervenes directly via terms involving ${\bf u}_0$), and centrifugal flattening (which acts through modification of the equilibrium state). To include the latter, we use a non-orthogonal, surface-matching coordinate system $(\zeta,\theta,\phi)$ \citep{Bonazzola1998} that has been employed by several authors in the calculation of stellar and planetary oscillation modes \citep{Lignieres2006,Reese2006,Reese2009,Reese2013,Reese2021,Ouazzani2012,Xu2017,Dewberry2021,Dewberry2022a,Dewberry2022b} and stellar structure \citep{Rieutord2016}. We then project the partial differential equations \eqref{eq:EoMl}-\eqref{eq:Poil} onto spherical harmonics, producing an infinite series of coupled sets of ordinary differential equations (ODEs) in the quasi-radial coordinate $\zeta$. Truncating this series at a maximum spherical harmonic degree $\ell_{\max}$, we solve the coupled ODEs simultaneously with a pseudospectral collocation method. We adopt a fiducial resolution of $100$ collocation points in the $\zeta-$direction, and increase $\ell_{\max}$ until the envelope of spectral coefficients becomes small (typically $\ell_{\max}\simeq16-100$).

\subsection{Tides}
This section lays out definitions, and introduces previous results from tidal theory that are relevant to the interpretation of our calculations.

\subsubsection{Tidal potential}
Assuming an orbital separation ${\bf d}$ sufficiently large for a tidal perturber to be treated as a point-mass $M'$, the tidal potential it imposes on the primary body can be written in terms of a multipole expansion as \citep{Jackson1962}
\begin{equation}\label{eq:U}
    U=-\frac{GM'}{a}
    \sum_{n=2}^\infty\sum_{m=-n}^n
    \left(\frac{4\pi}{2n+1}\right)
    \left(\frac{r}{a}\right)^n
    Y_n^{m*}(\theta',\phi')Y_n^m(\theta,\phi),
\end{equation}
where $a(t)=|{\bf d}|$, primes denote (time-dependant) satellite coordinates, and $Y_n^m$ are ortho-normalized spherical harmonics of degree $n$\footnote{The interplay between separate harmonics in the tidal potential and the response it induces motivates our use of both $\ell$ and $n$ for spherical harmonic degrees. We generally employ $n$ for degrees in the tidal potential that are summed over, reserving $\ell$ for the harmonic degree of interest in the induced response.} and azimuthal wavenumber $m$. Here we have neglected
the degree $n=0$ and $n=1$ terms in the expansion, which respectively have no effect and lead to basic Keplerian motion. 

Ignoring orbital eccentricity and inclination, this expansion can be written in the inertial frame as
\begin{equation}
    U=\sum_{n=2}^\infty\sum_{m=-n}^n
    U_{n m}r^n Y_n^m(\theta,\phi)
    \exp[-\text{i}\sigma_t t],
\end{equation}
where
\begin{align}
    U_{nm}&=-\left(\frac{GM'}{a^{n+1}}\right)\left(\frac{4\pi}{2n+1}\right)Y_n^{m*}(\pi/2,0),
\end{align}
and $\sigma_t$ is the inertial-frame tidal frequency. For this simplified case of a coplanar and circular orbit, $\sigma_t=m\Omega_o$, 
where $\Omega_o=[G(M+M')/a^3]^{1/2}$ is the mean motion of the perturber. Throughout, we adopt a nominal mass ratio of $q=M'/M=10^{-4}.$ This assumption only affects the results of our linear calculations by altering the relationship between $a$ and  $\Omega_o$.

\subsubsection{Potential Love numbers}\label{sec:love}
Fluid motions induced by the perturbing tidal potential will lead to an \textit{external} gravitational response $\Phi'$ that can in turn be expanded as
\begin{equation}\label{eq:PhEx}
    \Phi'=\sum_{n=2}^\infty\sum_{m=-n}^n
    \Phi'_{n m}r^{-(n+1)}
    Y_n^m\exp[-\text{i}\sigma_t t].
\end{equation}
For the linear tidal perturbation of an axisymmetric body, a coefficient $\Phi'_{\ell m}$ of a given degree $\ell$ and azimuthal wavenumber $m$ can be related to the coefficients $U_{nm}$ in the tidal potential via a linear relation involving potential ``Love numbers'' \citep{Ogilvie2013}:
\begin{equation}\label{eq:klnm}
    \Phi'_{\ell m}
    =\sum_{n=|m|}^\infty k_{\ell m}^n U_{nm}.
\end{equation}
A given $k^n_{\ell m}=k^n_{\ell m}(\sigma_t)$ thus specifies the amount to which a harmonic of degree $n$ in the tidal potential drives a gravitational response in degree $\ell$, at a given tidal frequency $\sigma_t$. 

In a spherically symmetric body $k_{\ell m}^n=0$ when $\ell\not=n$, but this is not true in general; in a rotationally flattened body harmonic coefficients of one degree in the induced tidal response cannot be solely attributed to coefficients of the same degree in the tidal potential. It is nevertheless still useful to consider the direct ratios 
\begin{equation}\label{eq:klm}
    k_{\ell m}=\frac{\Phi'_{\ell m}}{U_{\ell m}}
    =\sum_{n=|m|}^\infty k_{\ell m}^{n}
    \frac{U_{nm}}{U_{\ell m}},
\end{equation}
keeping in mind that these may not accurately reflect a causal relationship. In particular, in centrifugally flattened bodies the sectoral ($n=|m|$) part of the tidal potential can produce just as much of a tesseral ($n>|m|$) response as the corresponding tesseral part of the tidal potential. \cite{Dewberry2022a} showed that this sectoral driving of the tesseral response generically produces anomalously large $k_{\ell m}$ for all $\ell>|m|$,\footnote{See also \citet{Idini2022a}, who came to similar conclusions via a different approach.}
characterized by a strong dependence on the satellite separation $a$. 

Specifically $U_{nm}/U_{\ell m}\propto a^{\ell-n}$, so that for large $a$ and $\ell>|m|$ the term with $n=|m|$ dominates the sum over $n$ in Equation \eqref{eq:klm} if $k_{\ell m}^{|m|}$ is nonzero. For example, the values of $k_{42}$ reported by \citet{Wahl2020} for detailed Jupiter interior models perturbed by a static potential very closely follow the power law $k_{42}\propto a^2.$ Given a self-consistent tidal potential produced by satellites on Keplerian orbits, the spatial dependence $k_{\ell m}\propto a^{\ell -|m|}$ is equivalent to the frequency dependence 
\begin{equation}
    k_{\ell m}\propto \Omega_o^{-2(\ell-|m|)/3}
\end{equation} 
as $\Omega_o\rightarrow0$. The ratios $k_{\ell m}$ thus remain functions solely of frequency for a self-consistent tidal potential  \citep{Dewberry2022a}.

It is helpful to define a ``hydrostatic'' $k_{\ell m}^\text{hs}$ for rigidly rotating bodies via
\begin{equation}\label{eq:hs}
    k_{\ell m}^\text{hs}
    = \frac{1}{U_{\ell m}}
    \sum_{n=|m|}^\infty k_{\ell m}^n(\omega_t=0)U_{nm},
\end{equation}
where $\omega_t=\sigma_t-m\Omega$. These $k_{\ell m}^\text{hs}$ are not truly static, in that they depend on the satellite's motion (through frequency-dependent $U_{nm}/U_{\ell m}$), but they reproduce previous work employing static satellites at finite separation \citep[e.g.,][]{Wahl2017,Wahl2020,Nettelmann2019}. Consequently they can be used to isolate dynamical wave excitation.

\subsubsection{Tidal dissipation}\label{sec:diss}
The Love numbers $k_{\ell m}^n=k_{\ell m}^n(\sigma_t)$ are both frequency-dependant and complex-valued, their imaginary parts encoding a phase lag due to dissipation in the tidally perturbed body. For a viscous fluid, the time-averaged dissipation rate is the real part of \citep{Ogilvie2009}
\begin{equation}
    D_\nu 
    =-\frac{1}{2}\int_V
    {\bf v}^*\cdot(\nabla\cdot \delta{\bf T})
    \text{d}V.
\end{equation}
The energy and angular momentum transferred from the orbit to the primary due to the action of a given component of degree $\ell$ and order $m$ in the tidal potential---tidal power $P$ and torque $T$, respectively---can be computed from the imaginary parts of the Love numbers via \citep{Ogilvie2013}
\begin{align}
    P=\sigma_t\frac{(2\ell + 1)}{8\pi G}R_\text{eq}|U_{\ell m}|^2\text{Im}[k_{\ell m}^\ell]
    =(\sigma_t/m)T.
\end{align}
If the tidally perturbed planet or star rotates rigidly, the dissipation rate from a single component of the tidal potential can be related to the tidal power and torque via $D_\nu=P - \Omega T\propto\omega_t\text{Im}[k_{\ell m}^\ell]$. Our calculations verify this equality. Assuming the tidal distortion is dissipated, the requirement that $D_\nu$ be positive-definite then implies that the imaginary part of each $k_{\ell m}^\ell(\sigma_t)$ must have the same sign as $\omega_t$ \citep{Ogilvie2014}.

\begin{figure*}
    \centering
    \includegraphics[width=\textwidth]{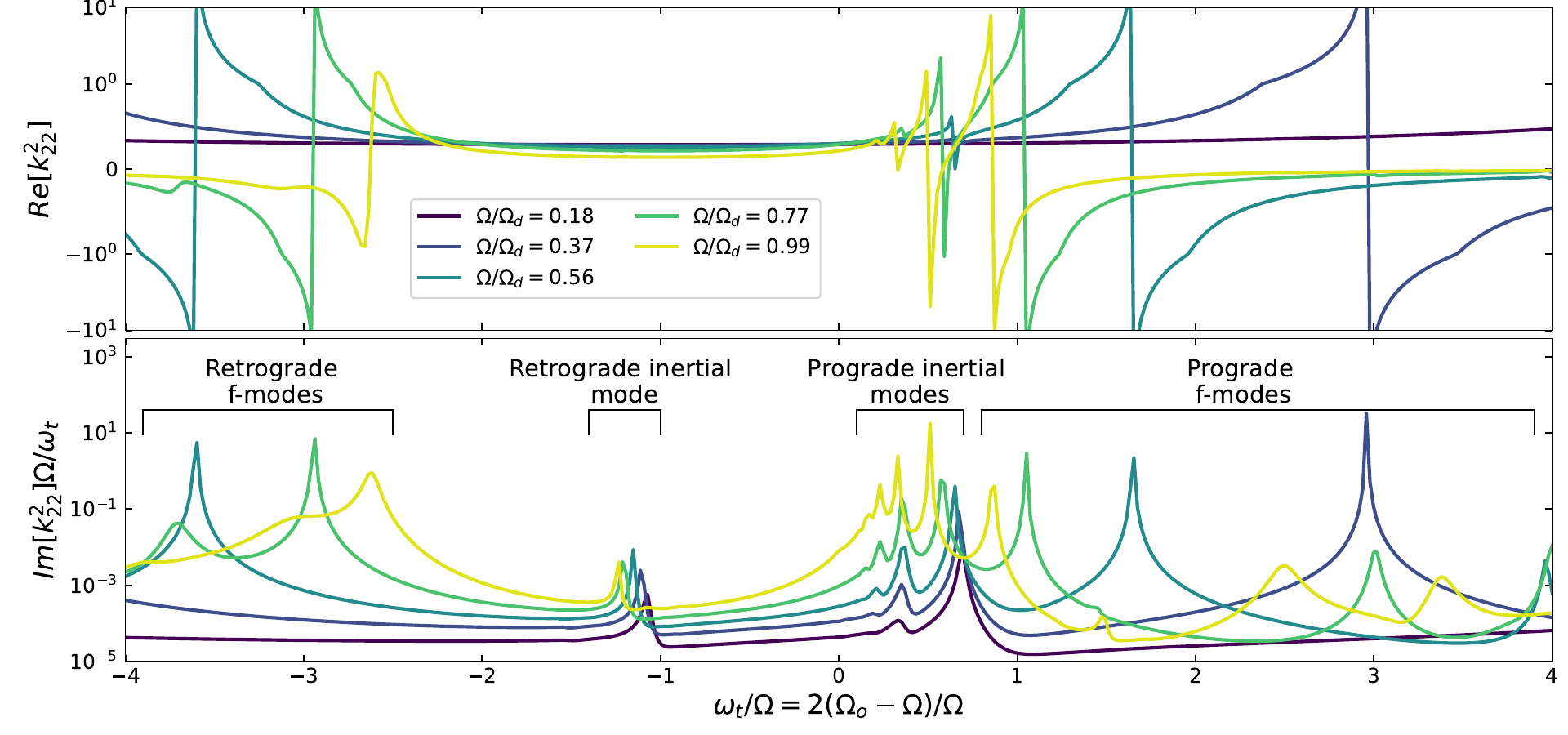}
    \caption{Real (top) and imaginary (bottom) parts of $k_{22}^2$ for rigidly rotating isentropic $\gamma=5/3$ polytropes with dynamic viscosity $\mu_v=10^{-5}$, plotted as a function of $\omega_t.$ From dark to light, line colors indicate increasingly rapid rotation. The top plot transitions from a log to linear scale at $Re[k_{22}^2]=1$. The peaks in $Im[k_{22}^2]\Omega/\omega_t$ correspond to tidal resonances with fundamental modes and inertial modes.}
    \label{fig:n15_k222}
\end{figure*}

\subsubsection{Modal expansion}\label{sec:modal}
We compute the tidal response both directly and through an expansion in the tidally driven oscillation modes of a rigidly rotating primary. The latter approach involves a phase space expansion of the form
\begin{equation}
    \left[ 
        \begin{matrix}
            \boldsymbol{\xi} \\
            \partial_t \boldsymbol{\xi}
        \end{matrix}
    \right]
    =\sum_\alpha c_\alpha(t)
    \left[
        \begin{matrix}
            \boldsymbol{\xi}_\alpha \\
            -\text{i}\omega_\alpha\boldsymbol{\xi}_\alpha
        \end{matrix}
    \right],
\end{equation}
where $\boldsymbol{\xi}_\alpha$ and $\omega_\alpha=\sigma_\alpha-m\Omega$ are the Lagrangian displacements and (rotating frame) frequencies of eigenmode solutions to Equations \eqref{eq:EoMl}-\eqref{eq:Poil} (in the absence of tidal forcing and viscosity), and $c_\alpha$ are tidally driven amplitudes. This sum over modes indexed by $\alpha$ includes all allowable $m,$ as well as complex conjugate ($\boldsymbol{\xi}_\alpha\mapsto\boldsymbol{\xi}_\alpha^*$) solutions.

In an inertial frame, the amplitude for a tidally driven oscillation mode of azimuthal wavenumber $m$ then satisfies \citep[e.g.,][]{Schenk2001,Lai2006}
\begin{equation}\label{eq:calph}
    \dot{c}_{\alpha}
    +\text{i}\sigma_{\alpha}c_{\alpha}
    =-\frac{\text{i}}{2\epsilon_{\alpha}}\exp[-\text{i}\sigma_tt]
    \sum_{n=|m|}^\infty U_{n m}Q_{n m}^\alpha,
\end{equation}
where 
\begin{align}
    \epsilon_\alpha 
    &=\omega_\alpha
    \langle \boldsymbol{\xi}_\alpha,\boldsymbol{\xi}_{\alpha}\rangle
    +\langle 
        \boldsymbol{\xi}_\alpha,
        \text{i}{\bf \Omega\times}\boldsymbol{\xi}_{\alpha}
    \rangle
\\\label{eq:Qlm}
    Q_{n m}^\alpha 
    &=\langle \boldsymbol{\xi}_\alpha,\nabla (r^n Y_n^m)\rangle
    =-\frac{(2n+1)}{4\pi}\Phi'_{nm,\alpha},
\end{align}
$ \langle \boldsymbol{\xi}_\alpha,\boldsymbol{\xi}_\beta\rangle
=\int_V\rho_0\boldsymbol{\xi}_\alpha^*\cdot\boldsymbol{\xi}_\beta\text{d}V$ defines an inner product, and $\Phi'_{nm,\alpha}$ is the contribution to the coefficient $\Phi_{nm}'$ in the expansion of Equation \eqref{eq:PhEx} that is attributable to the mode $\alpha$. These $Q_{nm}^\alpha$ coefficients are often referred to as overlap integrals. Steady-state solutions with $\dot{c}_\alpha=-\text{i}\sigma_tc_\alpha$ then satisfy
\begin{equation}
    c_\alpha
    =\frac{-\text{exp}[-\text{i}\sigma_tt]}
    {2\epsilon_{\alpha}(\sigma_\alpha-\sigma_t)}\sum_{n=|m|}^\infty U_{nm}Q_{n m}^\alpha
    \coloneqq \sum_{n=|m|}^\infty c_{\alpha}^n\text{exp}[-\text{i}\sigma_tt].
\end{equation}
Writing $\Phi'_{\ell m}=\sum_\alpha c_\alpha^\ell\Phi'_{\ell m,\alpha},$ Love numbers $k_{\ell m}^n$ can be computed by considering the effect of an isolated tidal potential of only one harmonic degree $n$:
\begin{equation}\label{eq:klmmode}
    k_{\ell m}^n
    =\frac{2\pi}{(2\ell+1)}
    \sum_{\alpha}
    \frac{Q_{\ell m}^\alpha Q_{nm}^\alpha}
    {\epsilon_{\alpha}(\sigma_\alpha-\sigma_t)}.
\end{equation}
Meanwhile, summing over $n$ in the full tidal potential provides \citep{Dewberry2022a}
\begin{equation}
    k_{\ell m}
    =\frac{2\pi}{(2\ell+1)}
    \sum_{\alpha}\sum_{n=|m|}^\infty
    \frac{Q_{\ell m}^\alpha Q_{nm}^\alpha}
    {\epsilon_{\alpha}(\sigma_\alpha-\sigma_t)}
    \left(\frac{U_{nm}}{U_{\ell m}}\right).
\end{equation}

\section{Fully isentropic and stratified polytropes}\label{sec:poly}
In this section we describe the results from tidal calculations for simple but very rapidly rotating polytropic models with equilibrium pressure and density related by $P_0\propto \rho_0^\gamma$. We consider two polytropic relations: $\gamma=5/3$ and $\gamma=3/2$. Together with a purely constant first adiabatic exponent $\Gamma_1=5/3$, $\gamma=5/3$ and $\gamma=3/2$ polytropes are neutrally and stably stratified throughout (respectively). The $\gamma=5/3$, neutrally stratified polytropes might be taken as reasonable models for fully convective compact objects or M-dwarfs. Meanwhile the $\gamma=3/2,$ stably stratified polytropes more closely approximate the interiors of main sequence stars.

Aside from their general applicability, the separate cases of fully isentropic and fully stratified stars provide a useful introduction to the partially stratified models of Jupiter considered in Section \ref{sec:Jup}. For both values of $\gamma$, we compute the $m=2$ tidal response for oblate models rotating at up to $99\%$ of the dynamical frequency $\Omega_d=(GM/R_\text{eq}^3)^{1/2}$. $\Omega_d$ provides a rough approximation to the critical ``mass-shedding'' limit at which the stars become unbound at the equator \citep[for $\gamma=5/3$ and $3/2,$ the mass-shedding limits are $\Omega\simeq1.02\Omega_d$ and $1.01\Omega_d$, respectively;][]{Dewberry2022b}.

\begin{figure*}
    \centering
    \includegraphics[width=\textwidth]{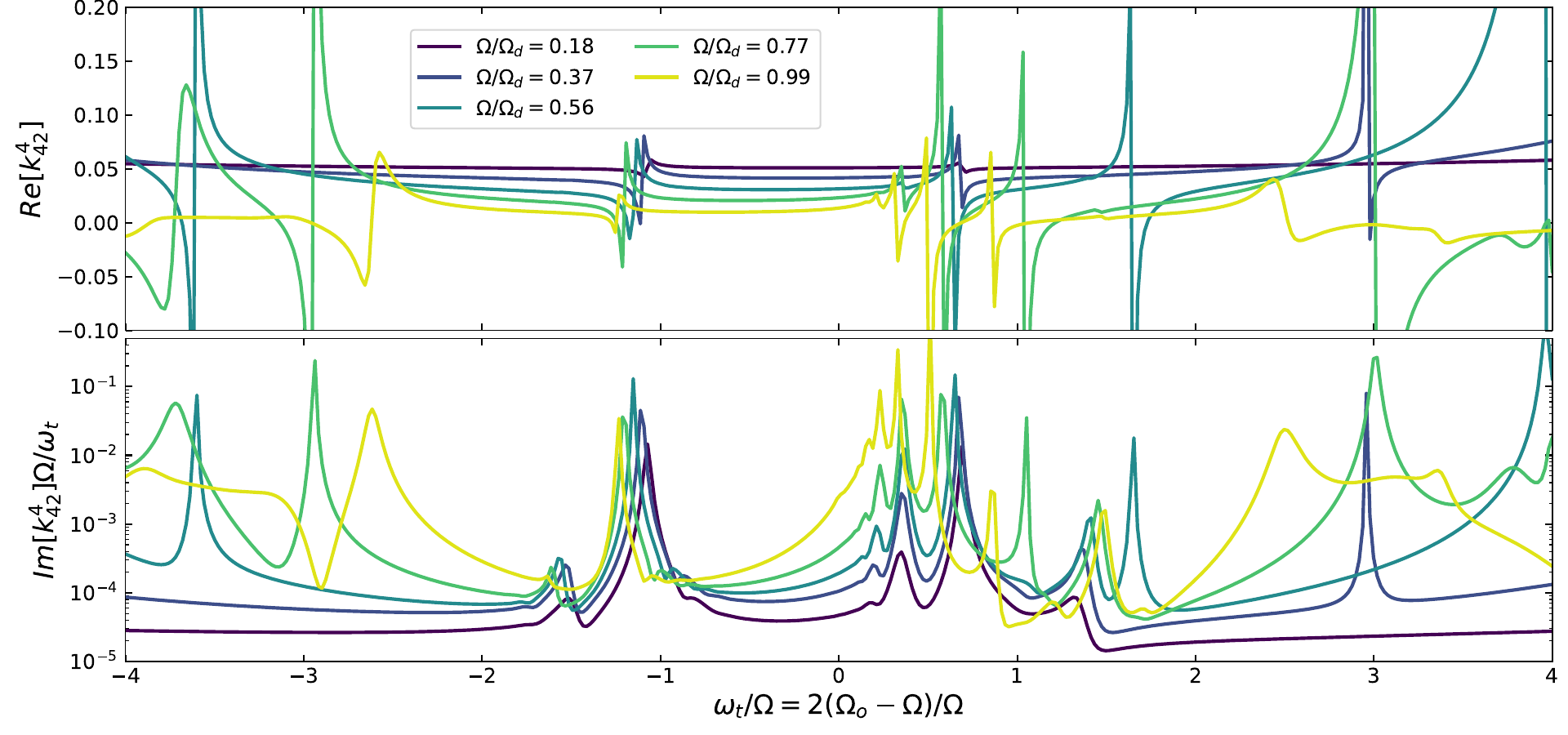}
    \caption{Same as Fig. \ref{fig:n15_k222}, but showing $k_{42}^4$ Love numbers computed for $\gamma=5/3$ polytropes perturbed by an isolated $n=4,m=2$ potential. The tesseral response of inertial modes plays a more significant role than in Fig. \ref{fig:n15_k222}.}
    \label{fig:n15_k424}
\end{figure*}

\subsection{Unstratified $\gamma=5/3$ polytropes}
The panels in Fig. \ref{fig:n15_k222} show the real (top) and imaginary (bottom) parts of Love numbers $k_{22}^2$ (Equation \ref{eq:klnm}) as a function of tidal frequency $\omega_t$, computed for isentropic $\gamma=5/3$ polytropes perturbed by a purely quadrupolar ($n=m=2$) tidal potential. The top panel employs a symmetric log-scale that transitions to linear at $Re[k_{22}^2]=1$. From dark to light, the line colors indicate polytropic models with increasingly rapid rotation. The calculations shown in Fig. \ref{fig:n15_k222} involved a constant dynamic viscosity $\mu_v=10^{-5}$ (in units with $G=M=R_\text{eq}=1$). This is large from an astrophysical perspective, but sufficiently small to reveal the important dynamical features of the model.

Resonances with internal oscillation modes produce sharp sign changes in $Re[k_{22}^2]$ and corresponding extrema in $Im[k_{22}^2]\Omega/\omega_t.$ Note that $Im[k_{22}^2]\Omega/\omega_t$ remains strictly positive, since sign$(Im[k_{\ell m}^n])=$sign$(\omega_t)$ \citep[see Section \ref{sec:diss}; ][]{Ogilvie2013}. The strong resonances at tidal frequencies $\omega_t/\Omega\lesssim -2$ and $\omega_t/\Omega\gtrsim 1$ correspond to retrograde and prograde fundamental modes (f-modes) with predominantly sectoral ($\ell\simeq m=2$) structure in their eigenfunctions. With faster and faster rotation, the natural frequencies of these oscillations become smaller in amplitude compared with the rotation rate \citep[e.g.,][]{Dewberry2022a}, and the resonances consequently move inward on an x-axis scaled by $\Omega$. For rotation rates $\Omega\gtrsim0.56$, higher degree ``tesseral'' ($\ell>m$) f-modes appear at higher frequencies in the bottom panel. They produce smaller resonances because of smaller spatial overlap with the $Y_2^2$ harmonic; in a non-rotating, spherically symmetric star the overlap integrals of tesseral f-modes with the sectoral tide vanish entirely.

\begin{figure}
    \centering
    \includegraphics[width=\columnwidth]{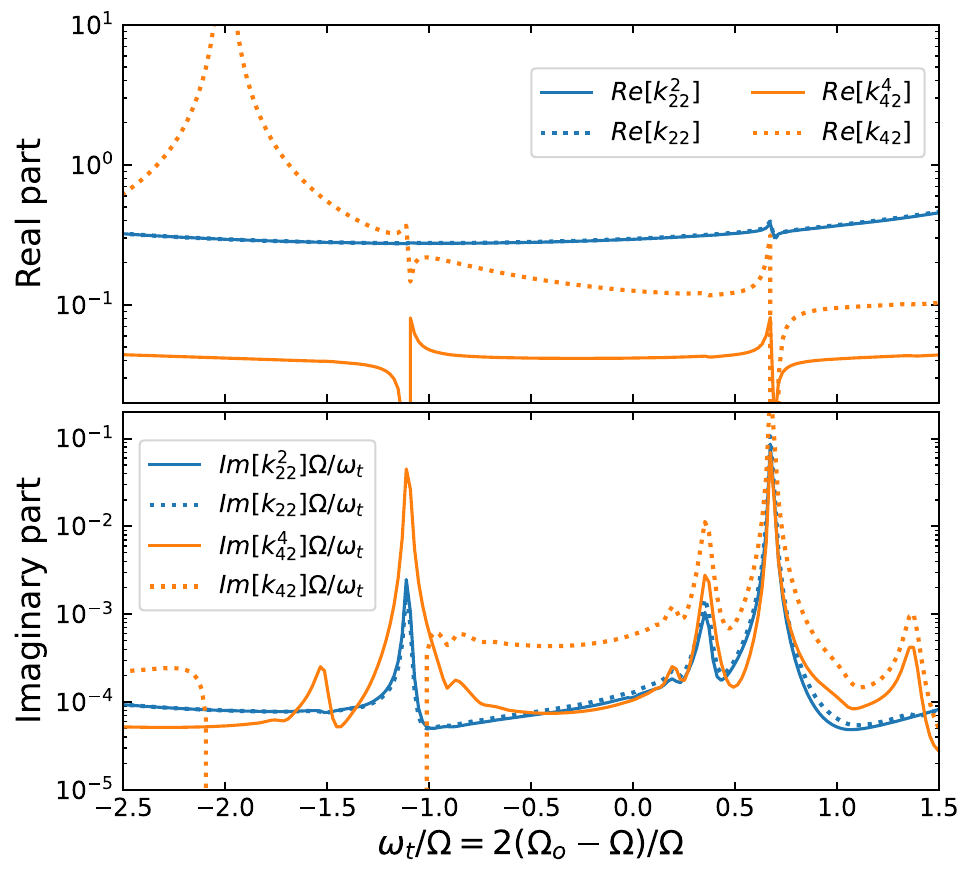}
    \caption{Plots comparing the real (top) and imaginary (bottom) parts of $k_{\ell m}^n$ (solid) against the direct ratios $k_{\ell m}$ (dotted) for a $\gamma=5/3$ polytrope with $\Omega/\Omega_d=0.37$. Although $k_{22}^2$ and $k_{22}$ agree, the $k_{42}^4$ and $k_{42}$ profiles deviate significantly due to centrifugal flattening \citep[see ][]{Dewberry2022a,Idini2022a}.}
    \label{fig:n15_klm_v_klmn}
\end{figure}

The bottom panel in Fig. \ref{fig:n15_k222} also illustrates some additional resonant peaks that remain fixed close to $\omega_t/\Omega\simeq-1.2$ and $\omega_t/\Omega\simeq0.6$ as the rotation rate increases. These resonances are produced by inertial modes \citep[e.g.,][]{Wu2005}, whose primary restoring force is the Coriolis. Inertial modes form a dense spectrum in the (rotating-frame) frequency range $-2\Omega<\omega<2\Omega,$ but only the longest wavelength modes couple strongly enough with the tidal potential to produce visible features in Fig. \ref{fig:n15_k222}. The solitary peaks near $\omega_t/\Omega\simeq-1.2$ correspond to the longest wavelength retrograde inertial mode, while the sequence of peaks with $\omega_t/\Omega\lesssim0.6$ are due to prograde inertial modes. The latter grow in amplitude with increasing rotation because of mixing (avoided crossing) with the prograde sectoral f-mode, as described in \citet{Dewberry2022a} for isentropic $\gamma=2$ polytropes.

Fig. \ref{fig:n15_k424} plots the real and imaginary parts of $k_{42}^4$ computed for the same $\gamma=5/3$ polytropes as shown in Fig. \ref{fig:n15_k222}. Since inertial oscillations generically couple more strongly to tesseral components of the tidal potential than sectoral \citep{Ogilvie2009,Ogilvie2013}, they feature more prominently in Fig. \ref{fig:n15_k424} than in Fig. \ref{fig:n15_k222}. 

\begin{figure*}
    \centering
    \includegraphics[width=\textwidth]{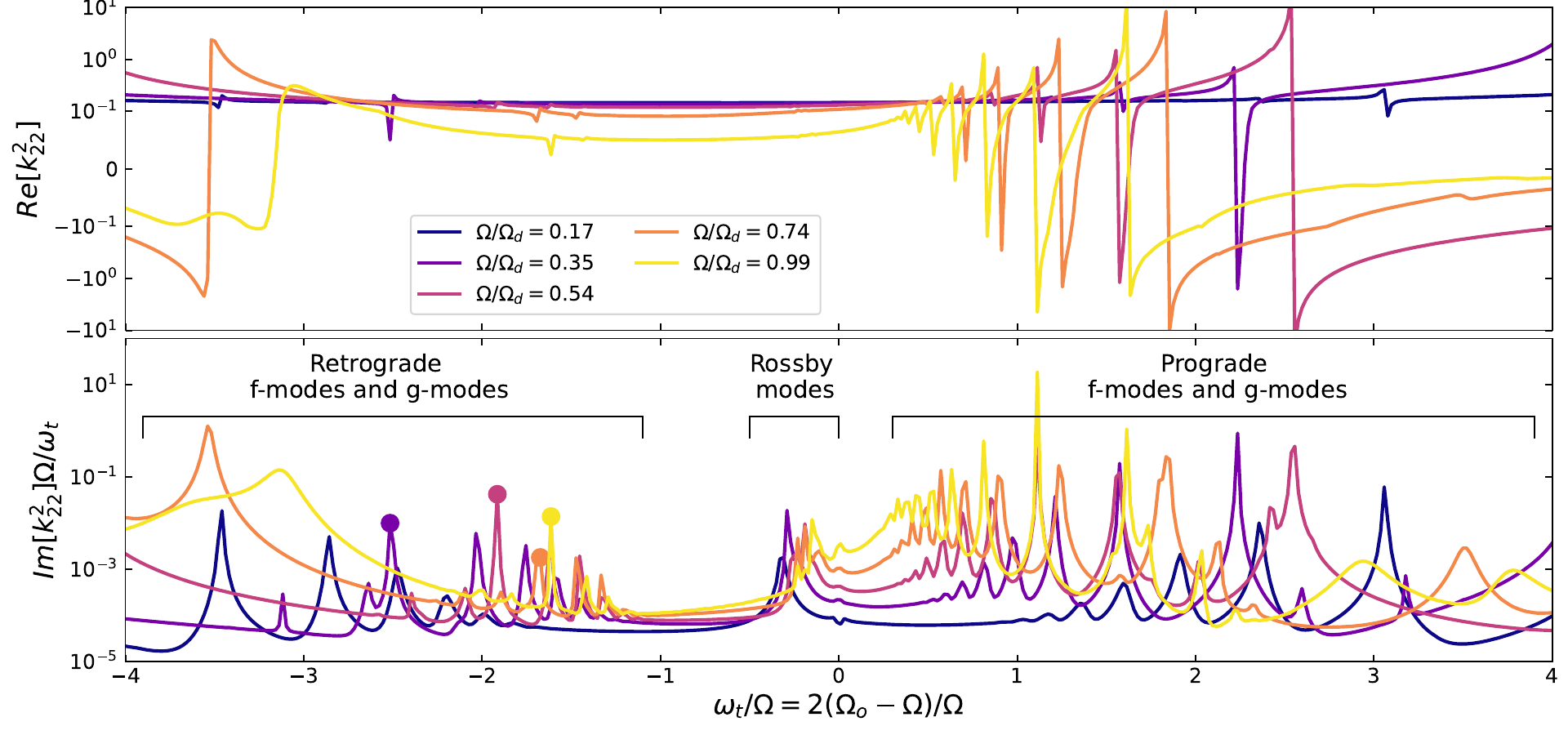}
    \caption{Same as Fig. \ref{fig:n15_k222}, but for $\gamma=3/2$ polytropes. Instead of inertial modes, the stable stratification in these stars supports gravito-inertial and Rossby modes. Filled circles indicate the frequencies corresponding to the cross-sections shown in Fig. \ref{fig:n2_ros}.}
    \label{fig:n2_k222}
\end{figure*}

The panels in Fig. \ref{fig:n15_klm_v_klmn} compare the real (top) and imaginary (bottom) parts of $k_{\ell m}^n$ (solid) and $k_{\ell m}$ (dotted) for the $\gamma=5/3$ polytrope with $\Omega/\Omega_d=0.37.$ We compute the latter by perturbing with a tidal potential including degrees $n=2-12$ (rather than isolated potentials of degree $n=2$ or $n=4$). $k_{22}\simeq k_{22}^2$, indicating (unsurprisingly) that the quadrupolar response of the star is dominated by the quadrupolar part of the tidal potential. On the other hand, Fig. \ref{fig:n15_klm_v_klmn} demonstrates dramatic disagreement between both the real and imaginary parts of $k_{42}^4$ and $k_{42}$. The apparent resonance in $Re[k_{42}]$ at $\omega_t/\Omega=-2$ has nothing to do with oscillations, instead reflecting the fact that as $\Omega_o\rightarrow0$ the tesseral response of centrifugally flattened bodies becomes dominated by the sectoral tide \citep[see Section \ref{sec:love};][]{Dewberry2022a,Idini2022a}. Additionally, $Im[k_{42}]/\omega_t$ becomes negative in the range $-2\Omega\lesssim\omega_t\lesssim-1$, disappearing from the log-scale of the plot in the bottom panel. Since $Im[k_{\ell m}^n]/\omega_t$ is strictly positive, negative values of $Im[k_{\ell m}]/\omega_t$ indicate frequency regimes where the induced tidal response in one harmonic is dominated by driving from a different harmonic in the tidal potential. The discrepancies between $k_{42}^4$ and $k_{42}$ are essential to our discussion of detectable resonances between Jupiter and its satellites in Section \ref{sec:Jup}.

\begin{figure*}
    \centering
    \includegraphics[width=\textwidth]{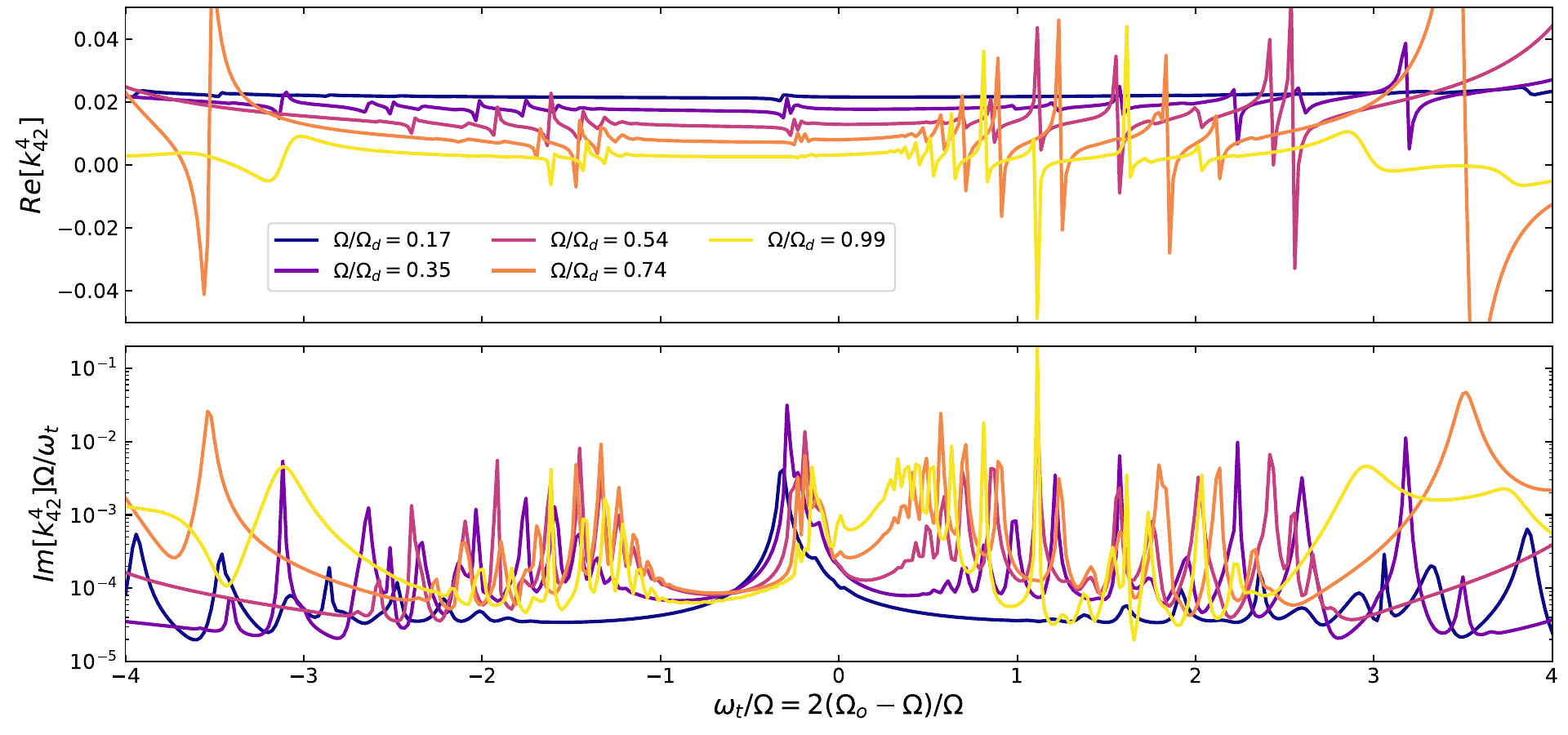}
    \caption{Same as Fig. \ref{fig:n15_k424}, but for $\gamma=3/2$ polytropes. Rotation causes g-mode eigenfunctions to overlap with multiple spherical harmonics, in turn leading to less regular sequences of peaks in $Im[k_{42}^4]$with increasingly rapid rotation.}
    \label{fig:n2_k424}
\end{figure*}
\begin{figure*}
    \centering
    \includegraphics[width=\textwidth]{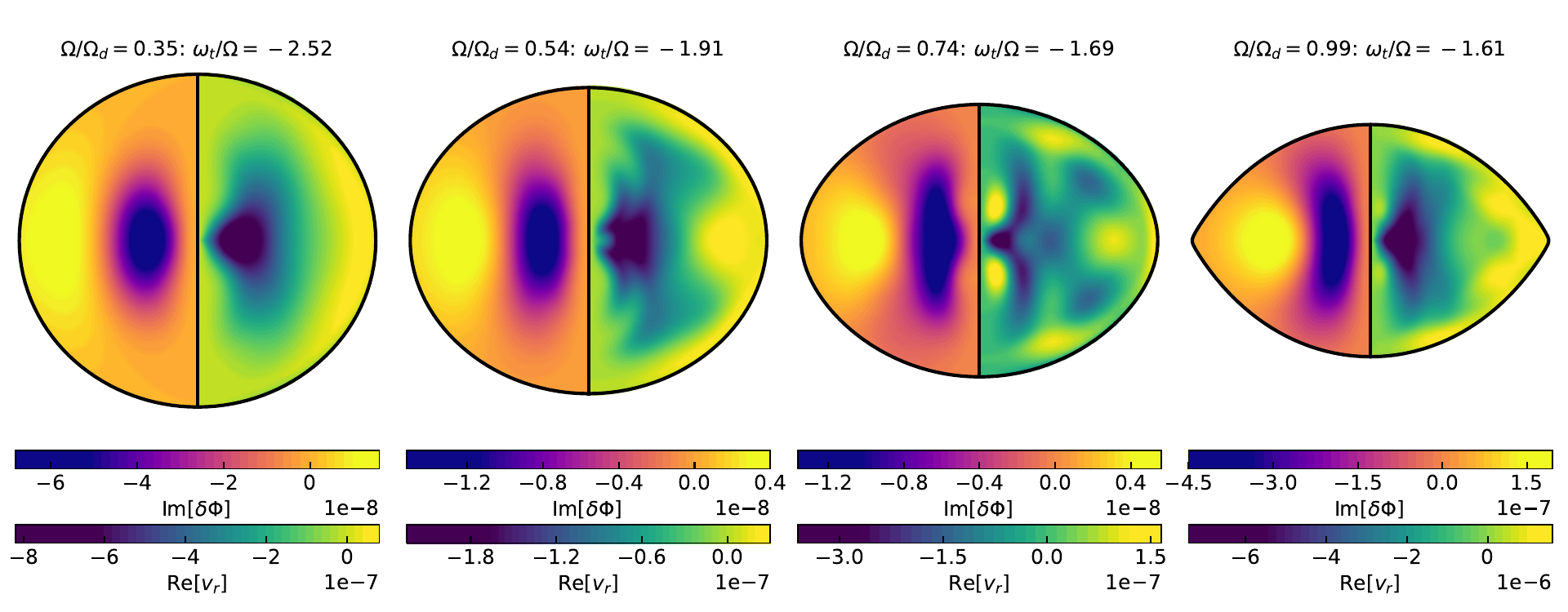}
    \caption{Cross-sections illustrating the gravitational (left) and radial velocity (right) perturbations induced in $\gamma=3/2$ polytropes at the tidal frequencies indicated by the filled circles in Fig. \ref{fig:n2_k222} (bottom). In rapid rotators, the angular structure of the tidal response is not well described by a single spherical harmonic.}
    \label{fig:n2_ros}
\end{figure*}

\subsection{Stratified $\gamma=3/2$ polytropes}
Figs. \ref{fig:n2_k222}-\ref{fig:n2_k424} are the same as \ref{fig:n15_k222}-\ref{fig:n15_k424}, but for stably stratified $\gamma=3/2$ polytropes. The resonances depicted in Figs. \ref{fig:n2_k222}-\ref{fig:n2_k424} consequently correspond to a different selection of internal oscillation modes. Along with f-modes, the peaks at higher frequencies $|\omega_t/\Omega|\gtrsim0.5$ are produced by gravito-inertial modes (g-modes), which are primarily restored by buoyancy. As shown in the bottom panel of \autoref{fig:n2_k222}, for a given rotation rate only a handful of long wavelength g-modes give rise to significant features in $Im[k_{22}^2]$ for this value of viscosity.  

Although the g-mode resonances are regularly spaced in $\omega_t/\Omega,$ their eigenfunctions can differ significantly from the g-modes of non-rotating stars. Rotation can confine g-mode eigenfunctions to the equator, and also mix together modes that in the limit $\Omega\rightarrow0$ have different harmonic degrees but nearly degenerate frequencies. The latter effect leads to ``rosette'' patterns in the oscillations' kinetic energy distributions \citep{Ballot2012,Takata2013,Dewberry2021}. The cross-sections (slices along the rotation axis) shown in Fig. \ref{fig:n2_ros} demonstrate the gravitational (left) and radial velocity (right) perturbations of the tidal response computed (using the full tidal potential) at the frequencies indicated by the filled circles in \ref{fig:n2_k222} (bottom). We plot the imaginary part of $\delta\Phi$ (and similarly the real part of $v_r$) because it better illustrates the structure of the resonant waves than the real part (with the phase chosen for the satellite, $Re[\delta\Phi]$ is dominated at most frequencies by the structure of non-resonantly driven f-modes). With increasingly rapid rotation, the induced wave patterns couple across a wide range of spherical harmonic degrees. 

Along with gravito-inertial and rosette waves at larger frequencies, Fig. \ref{fig:n2_k222} and \ref{fig:n2_k424} (bottom) reveal an additional family of resonant modes at $\omega_t/\Omega\simeq-0.3$. ``Rossby'' modes are purely retrograde oscillations restored by both the Coriolis and buoyancy forces \citep{Papaloizou1978,Townsend2003}. With relatively large tidal overlap integrals, Rossby modes may be important to tidal dissipation in super-synchronously rotating white dwarfs \citep{Fuller2014}. Recently, \citet{Papaloizou2023} have considered the role that resonant Rossby mode excitation may play in tidal interactions between exoplanets and their host stars.

\section{Tides in partially stratified bodies: application to  Jupiter}\label{sec:Jup}
In this section we consider the tidal response of Jupiter interior models that are simple but self-consistently flattened by rotation. Particular motivation for this application comes from the fact that Juno measurements of Jupiter's interaction with Io produce values of $k_{42}\simeq1.29$ \citep{Durante2020} that differ significantly from theoretical calculations of $k_{42}^\text{hs}\simeq1.74$ \citep{Wahl2017,Wahl2020,Nettelmann2019}. We find that dynamical tides are capable of reconciling this discrepancy, essentially due to dynamical driving of the tesseral response by the sectoral tide.

\begin{figure}
    \centering
    \includegraphics[width=\columnwidth]{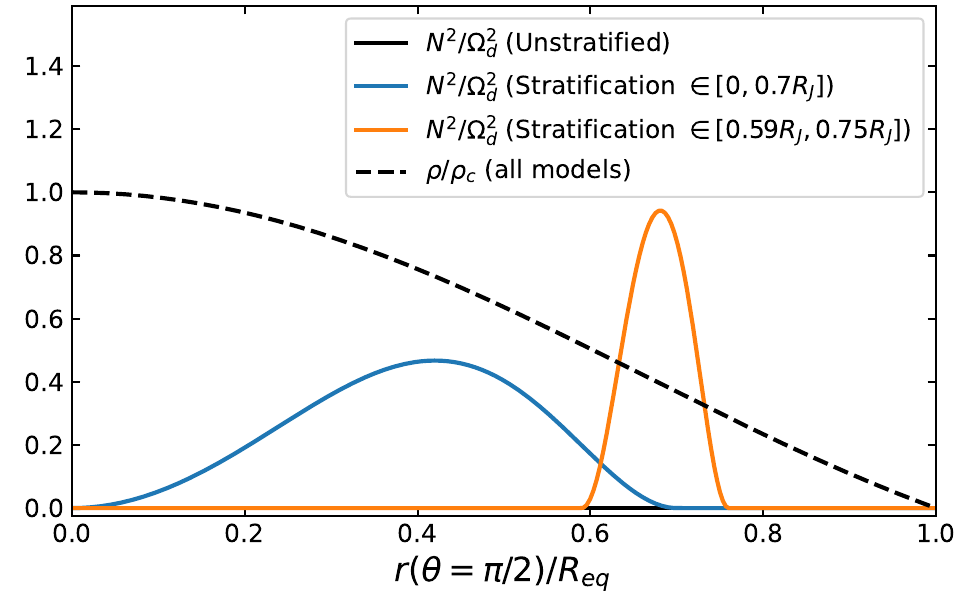}
    \caption{Equatorial profiles of the squared buoyancy frequency (solid) and density (dashed) for the Jupiter models considered in this paper. }
    \label{fig:jmod}
\end{figure}

\subsection{Interior models} 
We focus on $\gamma=2$ polytropes with Jupiter's bulk rotation rate $\Omega/\Omega_d\simeq 0.3$, and different profiles of stable stratification introduced via modification of $\Gamma_1.$ In particular, we assume the functional form 
\begin{equation}
    \Gamma_1(\zeta)
    =2 + \frac{A}{2}
    \left\{
        1 
        -\cos\left[ 
            2\pi\left(
            \frac{\zeta - \zeta_\text{i}}
            {\zeta_\text{o}-\zeta_\text{i}}
            \right)
        \right]
    \right\}.
\end{equation}
Here $A$ describes the amplitude of the deviation from isentropy, $\zeta$ is a dimensionless quasi-radial coordinate equal to one on the surface (see Appendix \ref{app:lin}), and $\zeta_\text{i},\zeta_\text{o}$ delimit the boundaries of a stably stratified region. Along with an isentropic model with $A=0,$ we consider partially stratified models characterized by $A=2,[\zeta_\text{i},\zeta_\text{o}]=[-0.7,0.7]$ and $A=0.5,[\zeta_\text{i},\zeta_\text{o}]=[0.59,0.76]$. Recent models involving a wide, stably stratified ``dilute'' core \citep{Wahl2017b,Militzer2022} motivate the former parameterization, while the latter produces a narrower band of stratification in the outer envelope \citep{Stevenson2022}. Choosing $\zeta_\text{i}<0$ for the dilute core model ensures even symmetry with respect to the origin. These profiles for stable stratification (shown in Fig. \ref{fig:jmod}) are intended only to capture the relevant wave dynamics, and not to serve as detailed interior models for Jupiter.

\begin{figure*}
    \centering
    \includegraphics[width=\textwidth]{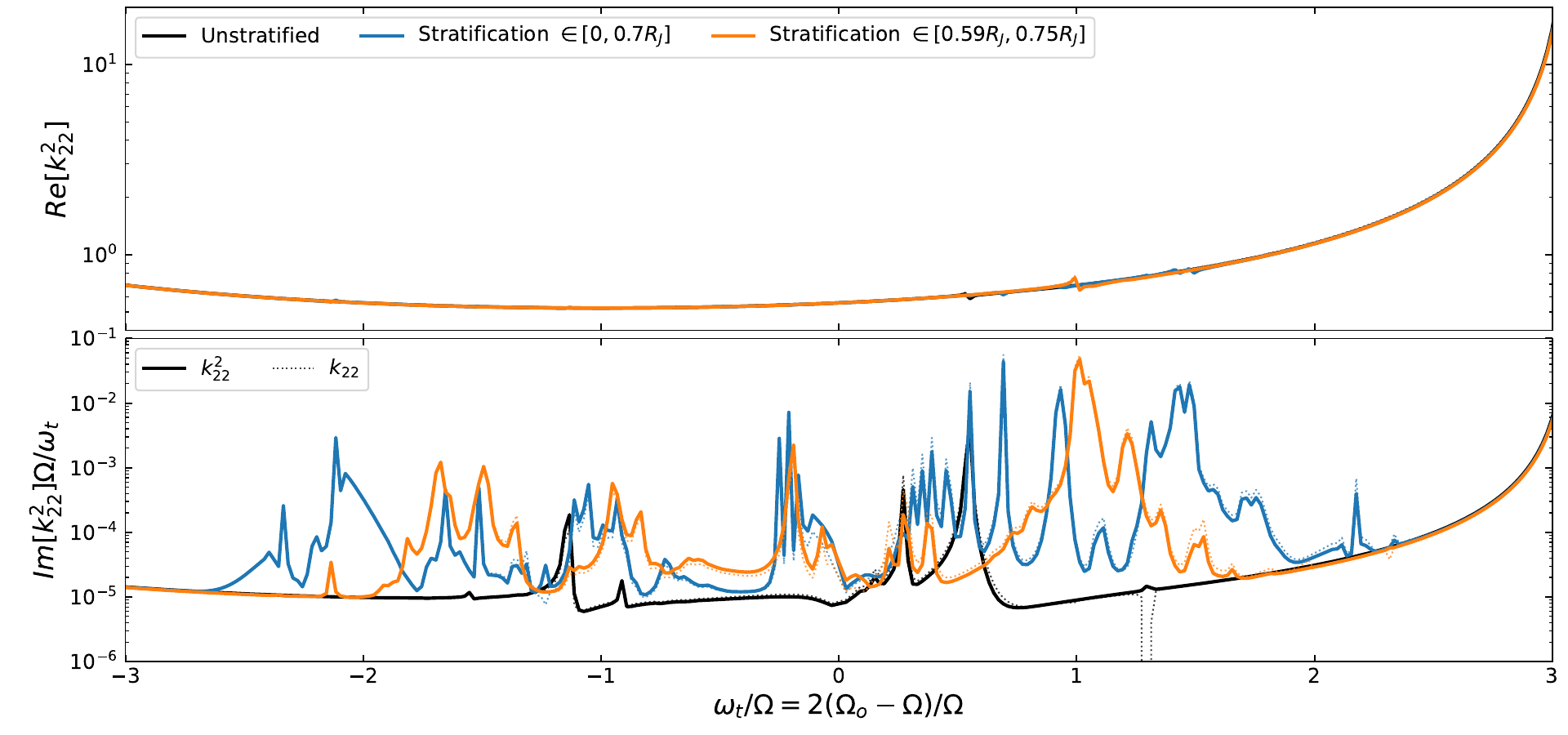}
    \caption{Profiles of $k_{22}^2$ like those of Fig. \ref{fig:n15_k222}, but for parameterized Jupiter interior models (see Fig. \ref{fig:jmod}) that are neutrally stratified (black), partially stratified with an expansive dilute core (blue), and partially stratified with a stable region in the envelope (orange). These calculations include a dynamic viscosity $\mu_v=10^{-6},$ and dotted lines indicate values of $k_{22}$ obtained for a tidal potential including degrees $n=2-12$.}
    \label{fig:jup_k222}
\end{figure*}
\begin{figure*}
    \centering
    \includegraphics[width=\textwidth]{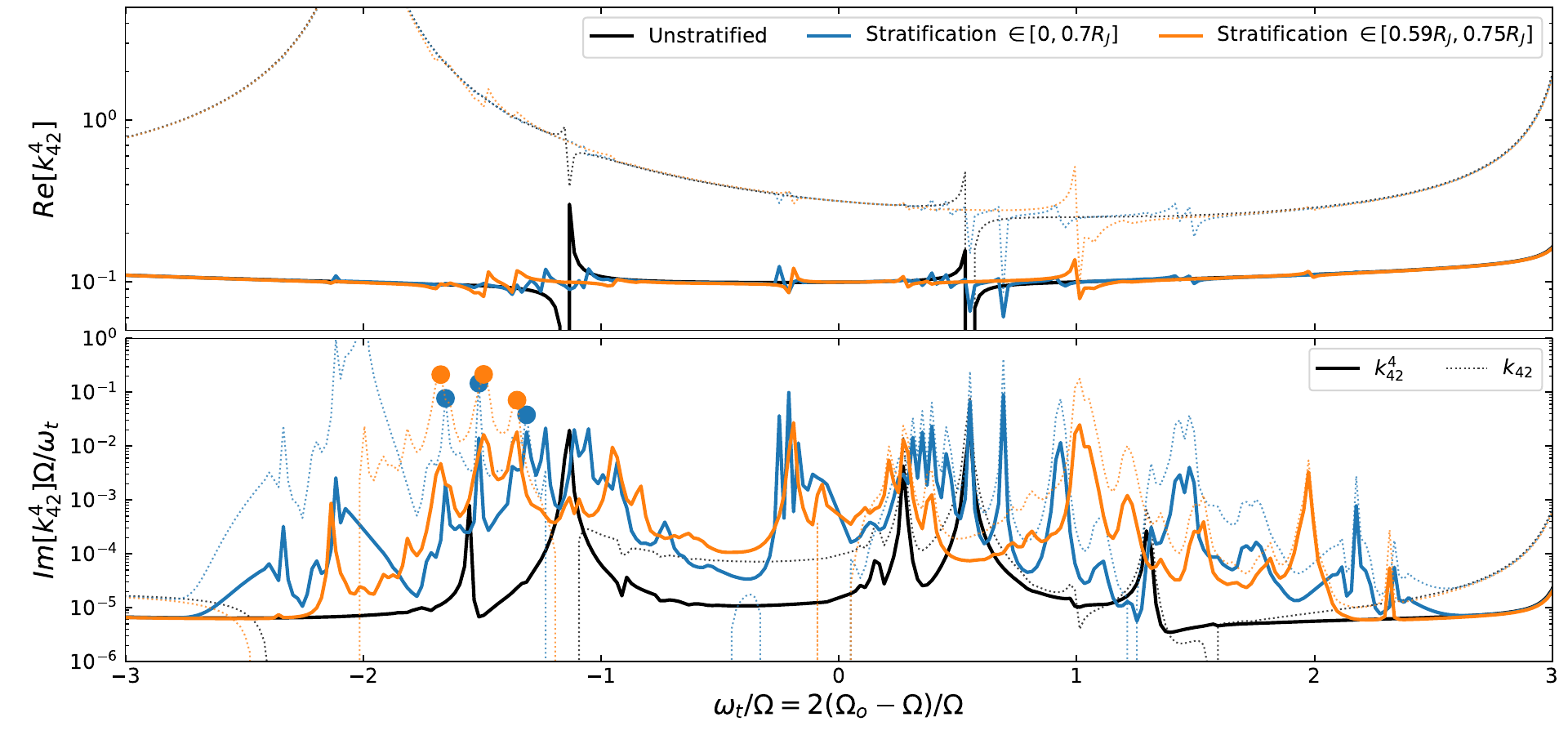}
    \caption{Same as Fig. \ref{fig:jup_k222}, but for $k_{42}^4$ (dotted lines show $k_{42}$). The filled circles denote frequencies corresponding to the cross-sections shown in Fig. \ref{fig:jupeaks}.}
    \label{fig:jup_k424}
\end{figure*}

\subsection{Tidal wave mixing}
Figs. \ref{fig:jup_k222}-\ref{fig:jup_k424} are similar to Figs. \ref{fig:n15_k222}-\ref{fig:n15_k424} and \ref{fig:n2_k222}-\ref{fig:n2_k424}. The solid lines plot real and imaginary parts of the Love numbers $k_{22}^2$ and $k_{42}^4$ describing $\ell=2$ and $\ell=4$ responses to isolated tidal potentials of the same degree, while the faint dashed lines show the ratios $k_{22}$ and $k_{42}$ computed by imposing a tidal potential including degrees $n=2-12.$ The black, blue and orange colors respectively indicate calculations for the completely isentropic, dilute core, and envelope stratification models introduced in the previous subsection. We adopt a dynamic viscosity of $\mu_v=10^{-6}$ for the calculations shown in these figures.

\begin{figure*}
    \centering
    \includegraphics[width=\textwidth]{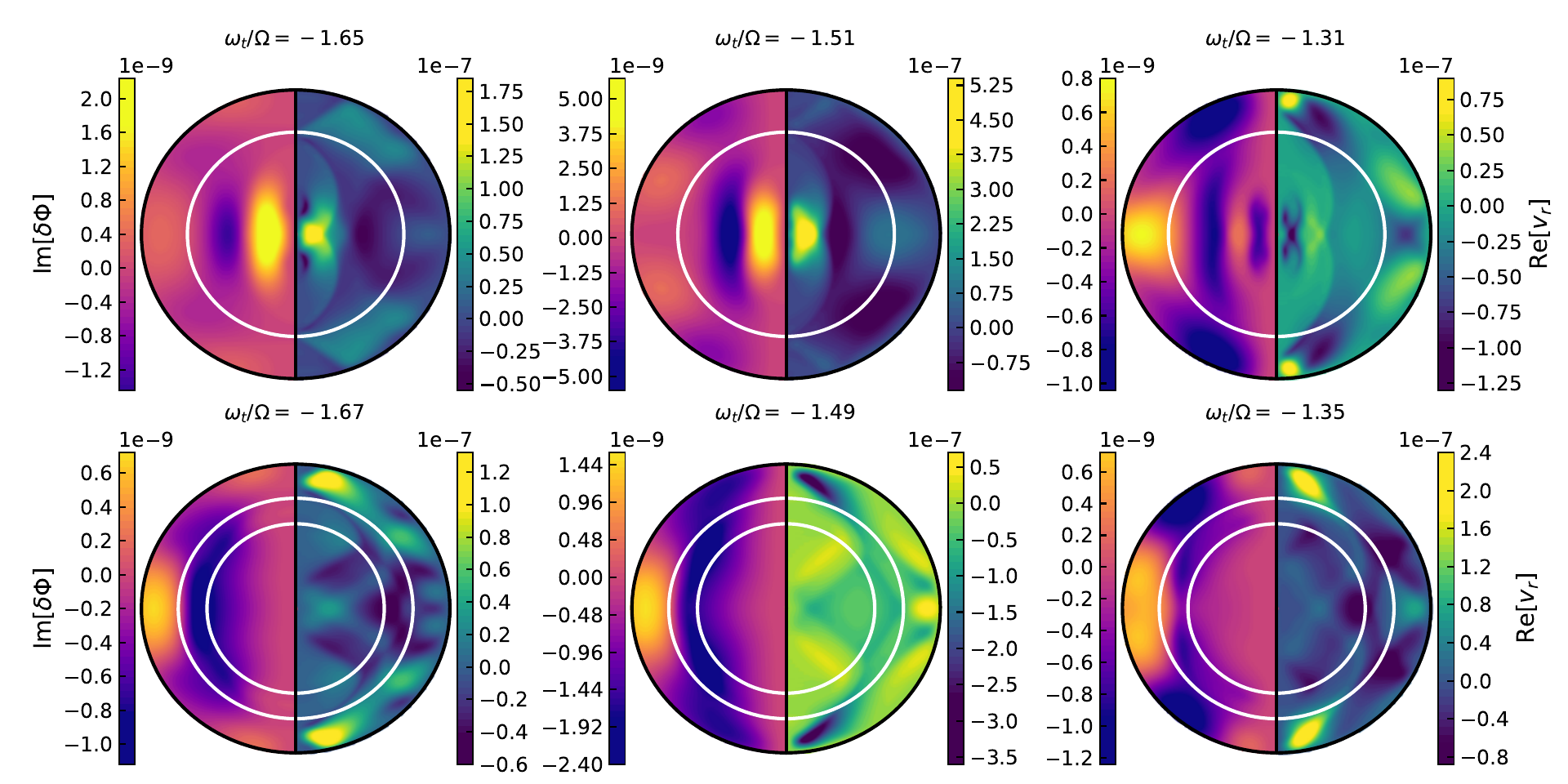}
    \caption{Cross-sections illustrating the gravitational (left) and radial velocity (right) perturbations induced in dilute core (top row) and envelope stratification (bottom row) models, at the tidal frequencies indicated by the filled circles in Fig. \ref{fig:jup_k424} (bottom). Interior white lines indicate boundaries between the convective and stably stratified regions. The partial stratification of these models leads to waves with gravito-inertial character in the stratified regions, and purely inertial character in the convective regions.}
    \label{fig:jupeaks}
\end{figure*}

The orange and blue curves demonstrate a more complicated spectrum of resonances than the polytropes considered in Section \ref{sec:poly}, owing to the fact that these Jupiter models possess \emph{both} convective and stably stratified regions. Inertial wave spectra are generically dense in frequency space \citep[e.g.,][]{Papaloizou1982}. All gravito-inertial modes with rotating-frame frequencies $\omega\in[-2\Omega,2\Omega]$ consequently have the capacity to mix with inertial waves in adjacent convective regions, so the waves forced by the tidal potential possess an inherently mixed character in the frequency range $\omega_t\in[-2\Omega,2\Omega]$. This is illustrated by the cross-sections in Fig. \ref{fig:jupeaks}, which (like Fig. \ref{fig:n2_ros}) show the gravitational and radial velocity perturbations induced by the full (multiple-degree) tidal potential at the resonant frequencies indicated by filled circles in Fig. \ref{fig:jup_k424} (bottom). The white lines indicate boundaries between the convective and stably stratified regions, with calculations for the dilute core and envelope stratification models shown in the top and bottom rows (respectively).

The cross-sections in Fig. \ref{fig:jupeaks} exhibit similar structure to the tidal waves computed by \citet{Lin2023} without the inclusion of centrifugal flattening. The gravito-inertial waves show the formation of rosette patterns, while the non-specular reflection of inertial waves off of the boundary between the stably stratified regions and the outer envelope leads to shorter wavelength beams of inertial waves propagating at an angle that depends on $\omega_t$ \citep{Ogilvie2009,Rieutord2010,Ogilvie2013,Lin2021,Lin2023}. 

Although the latter scattering to shorter wavelengths can enhance tidal dissipation at some frequencies, \citet{Lin2021} showed that the largest peaks in dissipation for isentropic planets with impermeable cores still correspond to underlying flows resembling the longest wavelength inertial modes of isentropic, coreless models. From the perspective of the modal expansion described in Section \ref{sec:modal}, the susceptibility of a given mixed mode $\alpha$ to excitation by a tidal potential of degree $n$ boils down to the requirement of a large ratio between overlap integrals $Q_{n  m}^\alpha$ and $\epsilon_\alpha$ coefficients. This requirement in turn filters for waves with some long-wavelength structure (for low $n$), regardless of whether those waves are also scattered to shorter wavelengths. Most importantly, the induced gravito-inertial and inertial waves of rotating models overlap with multiple spherical harmonic degrees, regardless of whether one or multiple harmonics are included in the perturbing tidal potential.

\begin{figure*}
    \centering
    \includegraphics[width=\textwidth]{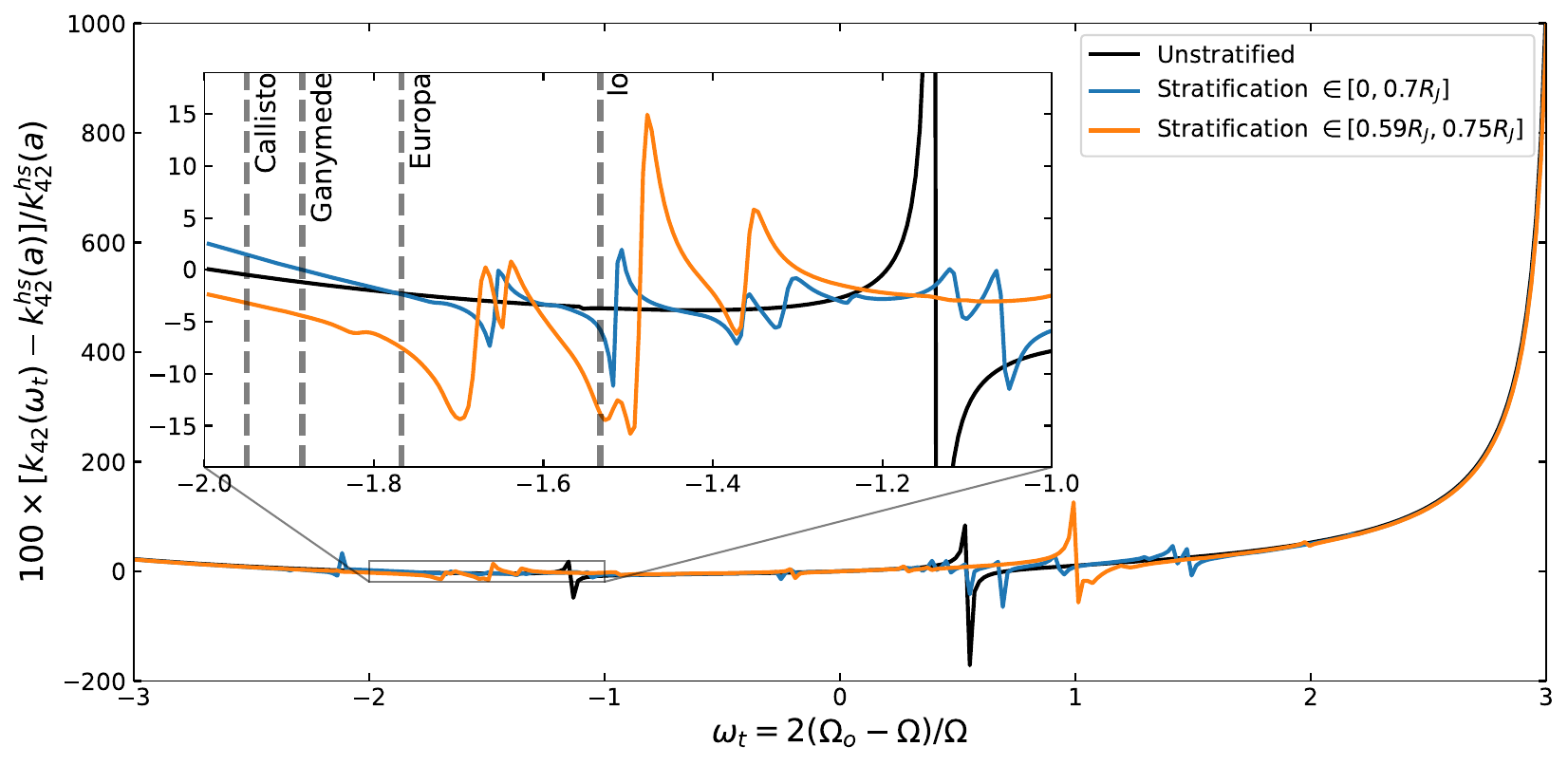}
    \caption{Per cent deviations in the real part of $k_{42}$ from the hydrostatic $k_{42}^\text{hs}$ (Equation \ref{eq:hs}), for the same models as Figs. \ref{fig:jup_k222}-\ref{fig:jup_k424}. As shown in the inset, resonances involving mixed waves at $\omega_t/\Omega\simeq-1.5$ (middle panels of Fig. \ref{fig:jupeaks}) lead to deviations of $\simeq-10\%$ to $-15\%$. Deviations of this magnitude are sufficient to reconcile observations and hydrostatic calculations to within $3\sigma$ \citep{Idini2022b}.}
    \label{fig:dk42}
\end{figure*}

\begin{figure*}
    \centering
    \includegraphics[width=\textwidth]{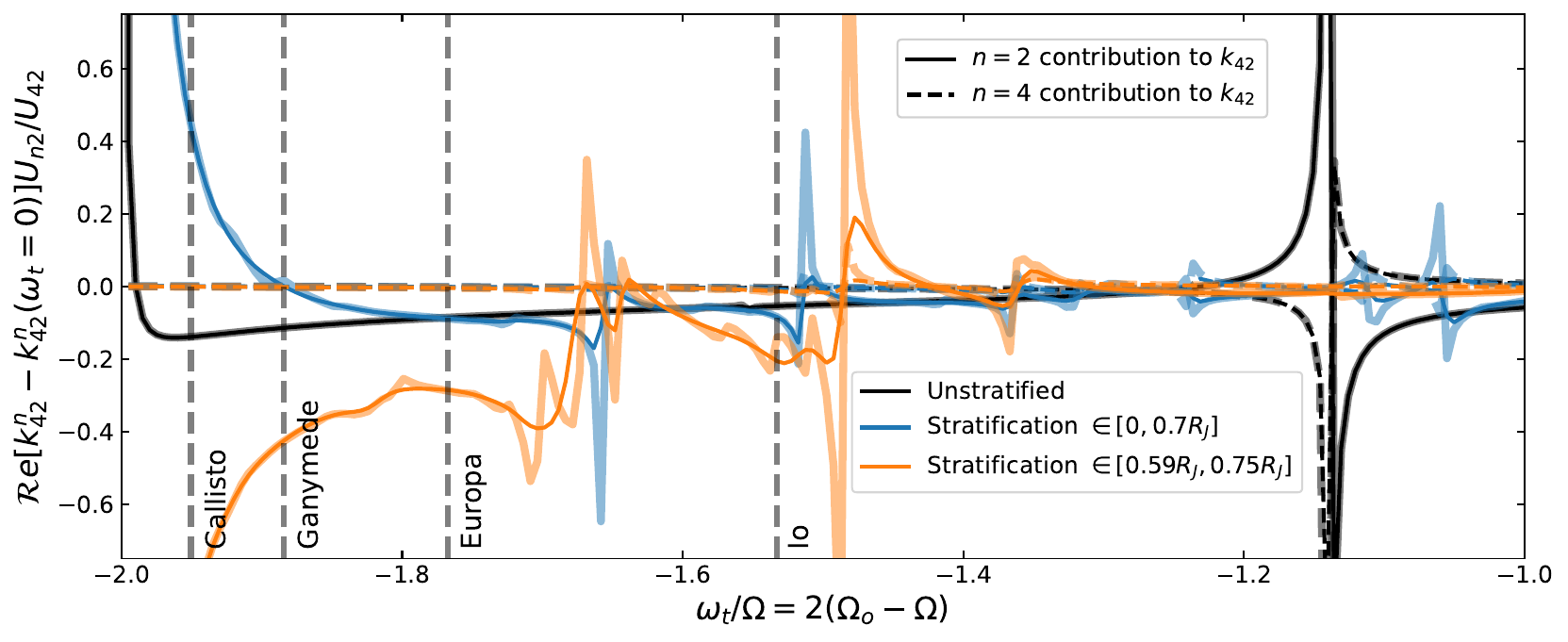}
    \caption{Curves describing the relative contributions of the degree $n=2$ (solid) and $n=4$ (dashed) parts of the tidal potential to the degree $\ell=4$ part of Jupiter's response. The $n=4$ contribution is inconsequential compared to that of $n=2$. Physically, this reflects the fact that the mixed waves of Fig. \ref{fig:jupeaks} overlap with both $n=2$ and $n=4$, but the quadrupolar forcing has a much larger amplitude at the tidal frequencies of the Galilean moons. The faint, thick lines show calculations with a lower viscosity ($\mu_v=10^{-7}$).}
    \label{fig:k42n}
\end{figure*}

\subsection{Jupiter's dynamical $k_{42}$}
Although our calculations produce qualitatively similar waves to those described by \citet{Lin2023}, with our inclusion of centrifugal flattening---excluded by both \citet{Idini2022b} and \citet{Lin2023}---we observe a much stronger dynamical impact on the tesseral ratios $k_{\ell m}$ with $\ell>m.$ Fig. \ref{fig:dk42} plots the per cent deviation of $Re[k_{42}]$ profiles computed for the three models shown in Figs. \ref{fig:jup_k222}-\ref{fig:jup_k424} from ``hydrostatic'' $Re[k_{42}^\text{hs}]$ (see Equation \ref{eq:hs}) values comparable with those computed by \citet{Wahl2017,Wahl2020} and \citet{Nettelmann2019}. We compute $k_{42}^\text{hs}$ profiles with the modal expansion described in Section \ref{sec:modal}, using oscillations calculated in the inviscid limit for the neutrally stratified model. Note that the real parts of the Love numbers for the isentropic and stratified Jupiter models considered here agree precisely in the limit $\omega_t\rightarrow0,$ so this approach yields a hydrostatic $k_{42}^\text{hs}$ that is relevant for all three.  

The inset highlights the frequency range relevant to Jupiter's Galilean moons, whose $m=2$ tidal frequencies are indicated by dashed grey lines. 
The solid black line again corresponds to the neutrally stratified model. Except for a feature near $\omega_t/\Omega\simeq-1.14$ corresponding to a resonance with the longest wavelength retrograde inertial mode, this model exhibits only gradual variation in $Re[k_{42}]$ due to the non-resonant influence of f-modes at much larger frequencies. As noted by \citet{Idini2022a}, this contributes a deviation of $\simeq-4\%$ from $k_{42}^\text{hs}$ at Io's frequency, which is insufficient to reconcile the tension between observations and hydrostatic calculations. 

On the other hand, the orange and blue curves (corresponding to the models with stratified regions in the outer envelope and core, respectively) exhibit much larger deviations. In particular, tidal wave excitation close to $\omega_t/\Omega\simeq-1.5$ (the middle cross-sections in Fig. \ref{fig:jupeaks}) induces deviations in $Re[k_{42}]$ of $10-15\%$ relative to the local $Re[k_{42}^\text{hs}]$. Such deviations are sufficient for agreement with the observed $k_{42}$ to within $3\sigma$ \citep{Idini2022a}. 

This result contrasts with the calculations of \citet{Lin2023}, which suggested that dynamically excited waves (with notably similar morphology to those shown in Fig. \ref{fig:jupeaks}) were incapable of significantly modifying the real part of $k_{42}$ (cf., their Fig. 6 and 8). The difference lies in our inclusion of rotational coupling across spherical harmonic degrees, as demonstrated by Fig. \ref{fig:k42n}. The curves in this figure plot $Re[k_{4 2}^n - k_{42}^n(\omega_t=0)]U_{n2}/U_{42}$ for $n=2$ (solid) and $n=4$ (dashed). This quantity describes the amount to which driving by the tidal potential of degree $n$ contributes to the response in degree $\ell=4.$ With the inclusion of centrifugal flattening, the Love numbers $k_{42}^2$ and $k_{42}^4$ are comparable for all of the waves shown in Fig. \ref{fig:jupeaks}. However, since $U_{22}\gg U_{42}$ as $\Omega_o\rightarrow0$ (i.e., as $\omega_t/\Omega\rightarrow-2$), the contribution from $n=2$ invariably dwarfs that from $n=4.$ Fig. \ref{fig:k42n} clearly demonstrates that wave coupling with the degree $n=2$ part of the tidal potential is the most important for the degree $\ell=4$ part of the tidal response.

We note that the dynamic viscosity used here ($\mu_v=10^{-6}$ in units with $G=M=R_\text{eq}=1$) is relatively large, ranging from Ekman numbers of $Ek=\mu_v/(\rho_0 \Omega R_\text{eq}^2)\simeq 10^{-6}$ at $r=0$ to $Ek\simeq10^{-3}$ close to the surface. The simplifying assumption of a constant dynamic (rather than kinematic) viscosity in particular leads to stronger damping in the outer envelope than may be realistic. However, we do not expect this to affect our main results: the faint, wider curves in Fig. \ref{fig:k42n} show calculations with a smaller $\mu_v=10^{-7}.$ Decreasing viscosity does little to affect the real parts of $k_{42}$ away from the strongest resonances, and only causes larger deviations close to resonance. Moreover, the results shown in Fig. \ref{fig:dk42} and Fig. \ref{fig:k42n} do not require moving particularly close to the strongest resonances (the widths of which vanish as $\mu_v\rightarrow0$).

The inherently mixed character of the resonantly driven waves shown in Fig. \ref{fig:jupeaks} means that labelling any of them as a g-mode of a particular harmonic degree is inappropriate. Nevertheless, we identify some of the resonances shown in Fig. \ref{fig:k42n} as involving what originate as $\ell=m=2$ g-modes in the non-rotating regime. We propose that resonances involving these oscillations provide as viable a candidate for the observed dynamical variation in Jupiter's $k_{42}$ as the $\ell=4$ g-modes considered by \citet{Idini2022b}; a resonantly driven wave need only have a non-negligible cross-product $Q_{42}Q_{22}$ of overlap integrals to contribute significantly to $k_{42}$ \citep{Dewberry2022a}.

\citet{Idini2022b} used an identification of the resonance with $\ell=4$ g-modes to infer an extended dilute core. Allowing for the possibility that Io may instead be in resonance with an $\ell\simeq2$ g-mode may lead to modification of these expectations for stable stratification in Jupiter. A more exhaustive survey of mixed-mode tidal resonances in a wider range of Jupiter interior models may therefore be worth pursuing.

\section{Conclusions}\label{sec:conc}
We have introduced a spectral numerical method for self-consistently computing the viscous tidal response of rapidly rotating, oblate planets and stars with arbitrary internal structures and rotation profiles. We have applied this method to fully isentropic (Figs. \ref{fig:n15_k222}-\ref{fig:n15_k424}) and fully stratified (Figs. \ref{fig:n2_k222}-\ref{fig:n2_ros}) polytropes with rigid rotation rates up to nearly the mass-shedding limit. We have also computed the tidal response for models of Jupiter's interior that include both stably stratified and convective regions (Figs. \ref{fig:jup_k222}-\ref{fig:jupeaks}). 

Contrary to recent work excluding centrifugal flattening \citep{Lin2023}, we find (Fig. \ref{fig:dk42}) that tidally excited oscillations in Jupiter are capable of reconciling a discrepancy between observed \citep{Durante2020} and predicted \citep{Nettelmann2019,Wahl2020} values of $k_{42}$ (the ratio between $\ell=4,m=2$ coefficients in multipole expansions of Jupiter's tidal response and Io's tidal potential). We find that in centrifugally flattened models, $\ell=2$ driving of mixed gravito-inertial and inertial waves contributes most significantly to Jupiter's $\ell=4$ response (Fig. \ref{fig:k42n}). Our results indicate that a wider set of internal oscillations than considered by \citet{Idini2022b} (in particular those originating as $\ell=2$ g-modes in the non-rotating regime) may serve as viable candidates for a Jupiter-Io resonance. Evaluating resonances with these additional oscillations in a range of realistic interior models may lead to modified constraints on Jupiter's stable stratification.

\section*{Acknowledgements}
I thank the reviewer for useful comments that significantly improved the quality of the paper. I also thank Jim Fuller, Dong Lai, and Yanqin Wu for helpful conversations. This work was supported by the Natural
Sciences and Engineering Research Council of Canada (NSERC), [funding
reference \#CITA 490888-16].

\section*{Data Availability}
The data underlying this work will be provided upon reasonable
request to the corresponding author



\bibliographystyle{mnras}
\bibliography{stratide}




\appendix
\onecolumn
\section{Linearized equations in non-orthogonal coordinates}\label{app:lin}
Equations \eqref{eq:EoMl}-\eqref{eq:Poil} can be written in tensor notation for an arbitrary curvilinear coordinate system as
\begin{align}\label{eq:EoMt}
    -\text{i}\sigma v^i
    +u_0^j\nabla_jv^i
    +v^j\nabla_j u_0^i
    +g^{ij}[
        (\partial_j +\partial_j\ln\rho_0)h
        -G_j\beta
        +\partial_j\delta\Phi
    ]
    -\frac{1}{\rho_0}\nabla_j \delta T^{ij}
    &=-g^{ij}\partial_j U,
\\\label{eq:Tt}
    \delta T^{ij}-\mu_v\left(
        g^{ik}\nabla_kv^j
        +g^{jk}\nabla_kv^i
        -\frac{2}{3}g^{ij}\nabla_kv^k
    \right)&=0,
\\\label{eq:Ctyt}
    -\text{i}\omega \beta+\frac{1}{J\rho_0}\partial_j(J\rho_0v^j)&=0,
\\\label{eq:TEt}
    -\text{i}\omega (h-c_A^2\beta)
    +v^j(G_j-c_A^2\partial_j\ln\rho_0)
    &=0,
\\\label{eq:Poit}
    4\pi G\rho_0\beta
    -\frac{1}{J}\partial_j(Jg^{jk}\partial_k\delta\Phi)
    &=0.
\end{align}
Here $\partial_i$ denotes partial differentiation with respect to the $i'$th curvilinear coordinate $x^i$, and upper (lower) indices denote contravariant (covariant) vector components associated with the expression of a vector in the covariant (contravariant) basis vectors ${\bf E}_i=\partial_i {\bf r}$ \ (${\bf E}^i=\nabla x^i$). Paired upper and lower indices denote summation, $g^{ij}={\bf E}^i\cdot{\bf E}^j$ is the inverse of the metric tensor $g_{ij}={\bf E}_i\cdot{\bf E}_j$, and $J=\sqrt{\det g_{ij}}$ is the Jacobian of the coordinate system. Lastly, $\nabla_i$ denotes covariant differentiation:
\begin{align}
    \nabla_k v^i
    &=\partial_kv^i
    +\Gamma^i_{jk}v^j,
\\
    \nabla_k\delta T^{ij}
    &=\partial_k \delta T^{ij}
    +\Gamma^i_{lk}\delta T^{lj}
    +\Gamma^j_{lk}\delta T^{il},
\end{align}
where $\Gamma^i_{jk}={\bf E}^i\cdot\partial_j {\bf E}_k$ are Christoffel symbols of the second kind. 

For a general mapping $r=r(\zeta,\theta)$ between spherical radius $r$ and a ``quasi-radial'' coordinate $\zeta$ \emph{defined} to be constant on the oblate (but still axisymmetric) surface $r_s=r_s(\theta)$, ${\bf E}_\zeta=\hat{\bf r}\partial_\zeta r,$ ${\bf E}_\theta=\hat{\bf r}\partial_\theta r + r\hat{\boldsymbol{\theta}}$, and ${\bf E}_\phi=r\sin\theta\hat{\boldsymbol{\phi}}$ \citep[e.g.,][]{Rieutord2016}. Our equilibrium velocity field is then simply ${\bf u}_0=\Omega {\bf E}_\phi$ (i.e., $u_0^\phi=\Omega$), and the metric tensor, specified by the line element $\text{d}s$ between two points, is given by 
\begin{equation}
    \text{d}s^2
    =g_{ij}\text{d}x^i\text{d}x^j
    =(\partial_\zeta r)^2\text{d}\zeta^2
    +\partial_\zeta r
    \partial_\theta r
    \text{d}\zeta\text{d}\theta
    +[(\partial_\theta r)^2 + r^2]\text{d}\theta^2
    +r^2\sin^2\theta \text{d}\phi^2.
\end{equation}
For such a coordinate system, $g^{ij}=g^{ji}$ and $\Gamma^i_{jk}=\Gamma^i_{kj}$. Additionally, 
\begin{equation}
    g^{\zeta\phi}
    =g^{\theta\phi}
    =\Gamma^\zeta_{\zeta\phi}
    =\Gamma^\zeta_{\theta\phi}
    =\Gamma^\theta_{\zeta\zeta}
    =\Gamma^\theta_{\phi\zeta}
    =\Gamma^\theta_{\phi\theta}
    =\Gamma^\phi_{\zeta\zeta}
    =\Gamma^\phi_{\theta\zeta}
    =\Gamma^\phi_{\theta\theta}
    =\Gamma^\phi_{\phi\phi} 
    =0.
\end{equation}
The viscous stress tensor involves six unique components. These can be eliminated from the equations by noting that covariant derivatives of $g^{ij}$ vanish, and that in $(\zeta,\theta,\phi)$ coordinates covariant derivatives commute (since the Riemann curvature tensor vanishes). For a constant dynamic viscosity $\mu_v$, $\nabla_j\delta T^{ij}$ can then be written as \citep{Hill2018}
\begin{equation}
    \nabla_j\delta T^{ij}
    =\mu_v \left[
            \frac{1}{J}\partial_j(Jg^{jk}\partial_kv^i)
            +2g^{jk}\Gamma^i_{kl}\partial_jv^l 
            +(g^{jk}\partial_l\Gamma^i_{kj})v^l
            +\frac{1}{3}g^{ik}\partial_k\mathcal{D}
    \right],
\end{equation}
where $\mathcal D=\nabla_iv^i$ is the velocity divergence. Retaining nonzero geometric factors, the separate components of the equations (expressed on the natural basis) then take the form
\begin{align}
\label{eq:EoMf1}
    \text{i}\omega v^\zeta
    -2\Omega\Gamma^\zeta_{\phi\phi}v^\phi
    +g^{\zeta j}[
        G_j\beta
        -(\partial_j +\partial_j\ln\rho_0)h
        -\partial_j\delta\Phi
    ]
    +\nu\left[
        \Delta_s v^\zeta
        +2g^{jk}\Gamma^\zeta_{kl}\partial_jv^l 
        +(g^{jk}\partial_l\Gamma^\zeta_{kj})v^l
        +\frac{1}{3}g^{\zeta k}\partial_k\mathcal{D}
    \right]
    &=g^{\zeta j}\partial_j U,
\\
\label{eq:EoMf2}
    \text{i}\omega v^\theta
    -2\Omega\Gamma^\theta_{\phi\phi}v^\phi
    +g^{\theta j}[
        G_j\beta
        -(\partial_j +\partial_j\ln\rho_0)h
        -\partial_j\delta\Phi
    ]
    +\nu\left[ 
        \Delta_s v^\theta
        +2g^{jk}\Gamma^\theta_{kl}\partial_jv^l 
        +(g^{jk}\partial_l\Gamma^\theta_{kj})v^l
        +\frac{1}{3}g^{\theta k}\partial_k\mathcal{D}
    \right]
    &=g^{\theta j}\partial_j U,
\\
\label{eq:EoMf3}
    \text{i}\omega v^\phi
    -(\partial_\zeta\Omega + 2\Omega\Gamma^\phi_{\zeta\phi})v^\zeta
    -(\partial_\theta\Omega + 2\Omega\Gamma^\phi_{\theta\phi})v^\theta
    -g^{\phi\phi}\partial_\phi (h+\delta\Phi)
    \hspace{15.5em}&\\\notag
    +\nu\left[ 
        \Delta_sv^\phi
        +2g^{jk}\Gamma^\phi_{kl}\partial_jv^l 
        +(g^{jk}\partial_l\Gamma^\phi_{kj})v^l
        +\frac{1}{3}g^{\phi\phi}\partial_\phi\mathcal{D}
    \right]
    &=g^{\phi\phi}\partial_\phi U,
\\
    \text{i}\omega \beta
    -(
        v^\zeta\partial_\zeta\ln\rho_0
        +v^\theta\partial_\theta\ln\rho_0
        +\mathcal{D}
    )
    &=0,
\\
    \text{i}\omega (h-c_A^2\beta)
    -(
        A_\zeta v^\zeta
        +A_\theta v^\theta
    )
    &=0,
\\
    \mathcal{D}
    -\left(
        \partial_\zeta 
        +\partial_\zeta\ln J
    \right)v^\zeta
    -\left(
        \partial_\theta 
        +\partial_\theta\ln J
    \right)v^\theta
    -\partial_\phi v^\phi
    &=0,
\\\label{eq:Poif}
    4\pi G\rho_0\beta-
    \left\{
        g^{\zeta\zeta}\partial^2_{\zeta\zeta}
        +[(\partial_j + \partial_j\ln J)g^{j\zeta}]\partial_\zeta
        +2g^{\zeta\theta}\partial^2_{\theta\zeta}
        +[(\partial_j + \partial_j\ln J)g^{j\theta}]\partial_\theta
        +g^{\theta\theta}\partial^2_{\theta\theta}
        +g^{\phi\phi}\partial^2_{\phi\phi}
    \right\}
    \Phi
    &=0,
\end{align}
where $\Delta_s=J^{-1}\partial_j(Jg^{jk}\partial_k\  {\bf\cdot}\  )$, $A_i=G_i-c_A^2\partial_i\ln\rho_0$, $\omega=\sigma - m\Omega,$ and $\nu=\mu_v/\rho_0$. 

\section{Numerical methods}\label{app:num}
To solve Equations \eqref{eq:EoMf1}-\eqref{eq:Poif}, we expand the perturbed variables in surface harmonics $Y_n^m$ that are normalized so that 
$\langle Y_\ell^m, Y_n^m\rangle
=\int_0^{2\pi}\int_0^\pi Y_\ell^m Y_n^m\sin\theta\text{d}\theta\text{d}\phi
=\delta_{\ell n}$, writing
\begin{align}\label{eq:exp1}
    v^\zeta 
    &=-\text{i}\sum_{n=|m|}^\infty Y_n^m(\theta,\phi)\tilde{a}^n(\zeta),
\\
    v^\theta
    &=-\frac{\text{i}}{\zeta}\sum_{n=|m|}^\infty 
    \left[
        \partial_\theta Y_n^m(\theta,\phi)\tilde{b}^n(\zeta)
        +\text{i}D_\phi Y_n^m(\theta,\phi)c^n(\zeta)
    \right],
\\
    v^\phi
    &=\frac{-1}{\zeta\sin\theta}\sum_{n=|m|}^\infty 
    \left[
        \text{i}D_\phi Y_n^m(\theta,\phi)\tilde{b}^n(\zeta)
        +\partial_\theta Y_n^m(\theta,\phi)c^n(\zeta)
    \right],
\\
    \beta 
    &=\sum_{n=|m|}^\infty Y_n^m(\theta,\phi)\beta^n(\zeta),
\\
    \mathcal{D}
    &=-\text{i}\sum_{n=|m|}^\infty Y_n^m(\theta,\phi)\tilde{\mathcal D}^n(\zeta),
\\
    h
    &=\sum_{n=|m|}^\infty Y_n^m(\theta,\phi)h^n(\zeta),
\\\label{eq:exp7}
    \delta\Phi
    &=\sum_{n=|m|}^\infty Y_n^m(\theta,\phi)\Phi^n(\zeta).
\end{align}
Here $D_\phi=(\sin\theta)^{-1}\partial_\phi$, and $\sim$'s indicate phase shifts introduced via factors of $\text{i}$. Note that the coefficients $\Phi^n$ differ from the $\Phi'_{n m,\alpha}$ discussed in Section \ref{sec:modal}, which are external multipole expansion coefficients in spherical rather than $(\zeta,\theta,\phi)$ coordinates. 
Writing $\mu=\cos\theta,$ $s=\sin\theta$ and substituting the spherical harmonic expansions (Equations \ref{eq:exp1}-\ref{eq:exp7}) into the linearized equations produces
\begin{align}\label{eq:EoMe1}
    \omega \zeta Y_n^m\tilde{a}^n
    +\frac{2\Omega}{s}
    \Gamma^\zeta_{\phi\phi}
    (
        \text{i}D_\phi Y_n^m\tilde{b}^n
        +\partial_\theta Y_n^mc^n
    )
    +\zeta g^{\zeta j}\left[
        G_jY_n^m\beta^n
        -(\partial_j +\partial_j\ln\rho_0)(Y_n^mh^n)
        -\partial_j(Y_n^m\Phi^n)
    \right]
    &
    \\\notag
    +\frac{\zeta }{\rho_0}\nabla_j\delta T^{ij}
    &=\zeta g^{\zeta j}\partial_j U,
\\\label{eq:EoMe2}
    \omega ( 
        \partial_\theta Y_n^m\tilde{b}^n  
        +\text{i}D_\phi Y_n^mc^n
    )
    +\frac{2\Omega}{s}
    \Gamma^\theta_{\phi\phi}
    (
        \text{i}D_\phi Y_n^m\tilde{b}^n 
        + \partial_\theta Y_n^mc^n)
    +\zeta g^{\theta j}[
        G_jY_n^m\beta^n
        -(\partial_j +\partial_j\ln\rho_0)(Y_n^mh^n)
        -\partial_j(Y_n^m\Phi^n)
    ]
    &\\\notag
    +\frac{\zeta }{\rho_0}\nabla_j\delta T^{\theta j}
    &=\zeta g^{\theta j}\partial_j U,
\\\label{eq:EoMe3}
    \omega (
        \text{i}D_\phi Y_n^m \tilde{b}^n 
        +\partial_\theta Y_n^mc^n
    )
    -\zeta s(\partial_\zeta\Omega + 2\Omega\Gamma^\phi_{\zeta\phi}) Y_n^m\tilde{a}^n
    -s(\partial_\theta\Omega + 2\Omega\Gamma^\phi_{\theta\phi})
    (
        \partial_\theta Y_n^m \tilde{b}^n
        + \text{i}D_\phi Y_n^mc^n
    )
    -\zeta sg^{\phi\phi}\text{i}\partial_\phi Y_n^m(h^n+\Phi^n)
    &\\\notag
    +\text{i}\frac{\zeta s}{\rho_0}\nabla_j \delta T^{\phi j}
    &=\zeta sg^{\phi\phi}\text{i}\partial_\phi U,
\\
    \omega \zeta Y_n^m\beta^n
    +\zeta \partial_\zeta\ln\rho_0\tilde{a}^nY_n^m
    +\partial_\theta\ln\rho_0
    (
        \partial_\theta Y_n^m \tilde{b}^n 
        +\text{i}D_\phi Y_n^m c^n
    )
    +\zeta Y_n^m \tilde{\mathcal{D}}^n
    &=0,
\\
    \omega\zeta Y_n^m(h^n - c_A^2\beta^n)
    +\zeta (G_\zeta - c_A^2\partial_\zeta\ln\rho_0)
    Y_n^m \tilde{a}^n
    +(G_\theta - c_A^2\partial_\theta\ln\rho_0)
    (
        \partial_\theta Y_n^m \tilde{b}^n 
        +\text{i}D_\phi Y_n^mc^n
    )
    &=0,
\\
    \zeta Y_n^m\tilde{\mathcal D}^n
    -\zeta Y_n^m\left(
        \partial_\zeta 
        +\partial_\zeta\ln J
    \right)\tilde{a}^n
    +[
        n(n+1)Y_n^m
        +(\mu/s - \partial_\theta \ln J)\partial_\theta Y_n^m
    ]\tilde{b}^n
    +(
        \mu/s - \partial_\theta \ln J
    )\text{i}D_\phi Y_n^mc^n
    &=0,
\\\label{eq:Poie}
    4\pi G\rho_0Y_n^m\beta^n
    -\Delta_s(Y_n^m\Phi^n)
    &=0,
\end{align}
where repeated indices $j$ and $n$ denote summation, and (assuming a constant dynamic viscosity)
\begin{align}
    \frac{\zeta }{\rho_0}\nabla_j \delta T^{\zeta j}
    &=
    -\text{i}\nu\Big\{ 
        \zeta \Delta_s (\tilde{a}^n Y_n^m) 
        +\zeta\left[
            2Y_n^mg^{\zeta k}\Gamma^\zeta_{k \zeta}\partial_\zeta
            +2g^{\theta k}\Gamma^\zeta_{k \zeta}\partial_\theta Y_n^m
            +(g^{jk}\partial_\zeta \Gamma^\zeta_{jk})Y_n^m
        \right]\tilde{a}^n
        \\\notag
        &\hspace{3em}
        +\left[
            2g^{\zeta k}\Gamma^\zeta_{k \theta}\partial_\theta Y_n^m(\partial_\zeta - 1/\zeta)
            +2g^{\theta k}\Gamma^\zeta_{k \theta}\partial^2_{\theta\theta} Y_n^m
            -2(m^2/s^2) g^{\phi\phi}\Gamma^\zeta_{\phi \phi}Y_n^m
            +(g^{jk}\partial_\theta \Gamma^\zeta_{jk})\partial_\theta Y_n^m
        \right]\tilde{b}^n
        \\\notag
        &\hspace{3em}
        -(m/s)\left[
            2g^{\zeta k}\Gamma^\zeta_{k \theta}Y_n^m
            (\partial_\zeta - 1/\zeta)
            +2g^{\theta k}\Gamma^\zeta_{k \theta}
            (\partial_\theta - \mu/s)Y_n^m
            -2g^{\phi\phi}\Gamma^\zeta_{\phi \phi}\partial_\theta Y_n^m
            +(g^{jk}\partial_\theta \Gamma^\zeta_{jk})Y_n^m
        \right]c^n
        \\\notag
        &\hspace{3em}
        +\frac{1}{3}\zeta (
            g^{\zeta \zeta}Y_n^m\partial_\zeta 
            +g^{\zeta \theta}\partial_\theta Y_n^m
        )\tilde{\mathcal D}^n
    \Big\},
\\
    \frac{\zeta }{\rho_0}\nabla_j\delta T^{\theta j}
    &=-\text{i}\nu\Big\{ 
        \zeta \left(
            2Y_n^mg^{\zeta k}\Gamma^\theta_{k \zeta}\partial_\zeta
            +2g^{\theta k}\Gamma^\theta_{k \zeta}\partial_\theta Y_n^m
            +g^{jk}\partial_\zeta \Gamma^\theta_{jk}Y_n^m
        \right)\tilde{a}^n
        +\zeta \Delta_s(\partial_\theta Y_n^m\tilde{b}^n/\zeta) 
        -m\zeta \Delta_s [Y_n^mc^n/(\zeta s)]
        \\\notag
        &\hspace{3em}
        +\left[
            2g^{\zeta k}\Gamma^\theta_{k \theta}\partial_\theta Y_n^m
            (\partial_\zeta - 1/\zeta)
            +2g^{\theta k}\Gamma^\theta_{k \theta}\partial^2_{\theta\theta} Y_n^m
            -2(m^2/s^2)g^{\phi\phi}\Gamma^\theta_{\phi \phi}Y_n^m
            +g^{jk}\partial_\theta \Gamma^\theta_{jk}\partial_\theta Y_n^m
        \right]\tilde{b}^n
        \\\notag
        &\hspace{3em}
        -(m/s)\left[
            2g^{\zeta k}\Gamma^\theta_{k \theta}Y_n^m(\partial_\zeta - 1/\zeta)
            +2g^{\theta k}\Gamma^\theta_{k \theta}(\partial_\theta - \mu/s)Y_n^m
            -2g^{\phi\phi}\Gamma^\theta_{\phi \phi}\partial_\theta Y_n^m
            +g^{jk}\partial_\theta \Gamma^\theta_{jk}Y_n^m
        \right]c^n
        \\\notag
        &\hspace{3em}
        +\frac{1}{3}\zeta (
            g^{\theta \zeta}Y_n^m\partial_\zeta 
            +g^{\theta \theta}\partial_\theta Y_n^m
        )\tilde{\mathcal D}^n
    \Big\},
\\
    \text{i}\frac{\zeta s}{\rho_0}\nabla_j \delta T^{\phi j}
    &=\text{i}\nu\Big\{
        2ms\zeta g^{\phi\phi}\Gamma^\phi_{\phi \zeta}Y_n^m\tilde{a}^n
        +ms\zeta \Delta_s[Y_n^m\tilde{b}^n/(s^2\zeta)]
        -s\zeta \Delta_s[\partial_\theta Y_n^mc^n/(s\zeta)]
        \\\notag
        &\hspace{3em}
        +2(m/s)\left[
            g^{\zeta k}\Gamma^\phi_{k \phi}Y_n^m
            (\partial_\zeta - 1/\zeta)
            +g^{\theta k}\Gamma^\phi_{k \phi}(
                \partial_\theta 
                - 2\mu/s
            )Y_n^m
            +s^2g^{\phi\phi}\Gamma^\phi_{\phi \theta}\partial_\theta Y_n^m
        \right]\tilde{b}^n
        \\\notag
        &\hspace{3em}
        -2\left[
            \partial_\theta Y_n^m g^{\zeta k}\Gamma^\phi_{k \phi}
            (\partial_\zeta - 1/\zeta)
            +g^{\theta k}\Gamma^\phi_{k \phi}
            (
                \partial_\theta 
                - \mu/s
            )\partial_\theta Y_n^m
            +m^2g^{\phi\phi}\Gamma^\phi_{\phi \theta}Y_n^m
        \right]c^n
        -\frac{1}{3}s^2\zeta g^{\phi \phi}\text{i}D_\phi Y_n^m \tilde{\mathcal D}^n
    \Big\}.
\end{align}
Note that all but the dissipative terms in Equations \eqref{eq:EoMe1}-\eqref{eq:Poie} involve purely real coefficients. We reduce dimensionality through spherical harmonic projection, first taking the inner product of the $\zeta$-component of the equation of motion with $Y_\ell^m$ for an arbitrary degree $\ell.$ For the angular components of the equation of motion we follow \cite{Reese2006}, operating with 
$\langle \partial_\theta Y_\ell^m,$\eqref{eq:EoMe2}$\rangle
-\langle\text{i}D_\phi Y_\ell^m,$\eqref{eq:EoMe3}$\rangle$
and 
$\langle \text{i}D_\phi Y_\ell^m,$\eqref{eq:EoMe2}$\rangle
-\langle \partial_\theta Y_\ell^m,$\eqref{eq:EoMe3}$\rangle$.
We finally take the inner product of the continuity, energy, divergence, and Poisson equations with $Y_\ell^m$. When $\Omega=0$, the projected equations separate to produce independent sets of ordinary differential equations (one set for each degree $\ell$ and azimuthal wavenmber $m$) in radius. In a rotating planet or star, the Coriolis force and centrifugal flattening couple the equations of one degree to another. 

We solve the coupled series of ODEs simultaneously using pseudospectral collocation, computing solutions $X=[X^{|m|},X^{|m|+1},...,X^{\ell_{\max}}]^T$, where $X^n=[\tilde{a}^n,\tilde{b}^n,c^n,\beta^n,h^n,\tilde{\mathcal D}^n,\Phi^n],$ to the boundary value and eigenvalue problems posed by Equations \eqref{eq:EoMf1}-\eqref{eq:Poif}. Numerical tractability requires truncation at a maximum degree $\ell_{\max}$. For all of the calculations in this paper, we set $\ell_\text{max}$ so that $\text{max}|X^n|$ for $n=\ell_\text{max}$ is at least $1000$ times smaller than for any other degree $n,$ and adopt a fiducial resolution of $N_\zeta=100$. We find little difference with increasing $N_\zeta$ for the viscosities used, except near $\omega_t=0$ for the $\gamma=3/2$ polytropes (note the small discontinuities near $\omega_t=0$ in the bottom panels Figs. \ref{fig:n2_k222} and \ref{fig:n2_k424}). Inaccuracy at this frequency should have no effect on the rest of our results, and so we deem $N_\zeta=100$ to be sufficient.

We use boundary bordering to enforce boundary conditions for each $X^n$: at the origin, we enforce regularity in all of the perturbed variables. We additionally enforce the continuity of the gravitational potential and its gradient at the perturbed surface \citep{Reese2013}, and match $\Phi$ to a potential in the external vacuum that vanishes at infinity. We finally require that the normal and tangential stresses vanish at the surface. The total stress at the perturbed surface is defined by the tensor $\mathcal{T}^{ij}=-\Delta Pg^{ij} + \delta T^{ij}$, where $\Delta P=\delta P + \boldsymbol{\xi}\cdot\nabla P_0$ is the Lagrangian pressure perturbation, and $\boldsymbol{\xi}$ is the Lagrangian displacement (related to the Eulerian velocity perturbation by 
${\bf v}=\partial_t \boldsymbol{\xi}
+{\bf u}_0\cdot\nabla\boldsymbol{\xi}
-\boldsymbol{\xi}\cdot\nabla{\bf u}_0$). Note that since the un-perturbed oblate surface is defined by the equation $S=\zeta-1=0,$ its surface normal vector is given by ${\bf n}=\nabla S/|\nabla S|=(g^{\zeta\zeta})^{-1/2}{\bf E}^\zeta\coloneqq  n_\zeta {\bf E}^\zeta$. The condition that the stress vanish at the surface can then be written as $\mathcal T^{ij}n_j=\mathcal T^{i\zeta}n_\zeta=0$, or 
\begin{align}
    -(\delta P + \boldsymbol{\xi}\cdot\nabla P_0)g^{\zeta\zeta} + \delta T^{\zeta\zeta}&=0,
\\
    -(\delta P + \boldsymbol{\xi}\cdot\nabla P_0)g^{\zeta\theta} + \delta T^{\zeta\theta}&=0,
\\
    \delta T^{\zeta\phi}&=0.
\end{align}
In order to retain separability in the limit $\Omega\rightarrow0,$ we project the latter two boundary conditions following a similar procedure to the angular components of the equation of motion.


\bsp	
\label{lastpage}
\end{document}